\documentclass[fleqn,usenatbib,twocolumn]{mnras}

\usepackage{newtxtext,newtxmath}

\usepackage[T1]{fontenc}
\usepackage{ae,aecompl}

\usepackage{graphicx}	
\usepackage{amsmath}	
\usepackage{amssymb}	
\usepackage{times}
\usepackage{footnote}
\usepackage{paralist}
\usepackage{subfigure}
\usepackage{siunitx}
\DeclareSIUnit\year{year}
\usepackage{booktabs}
\usepackage[dvipsnames]{xcolor}
\usepackage[normalem]{ulem}

\newcommand{\bs}{\boldsymbol}
\newcommand{\percent}{\mathrm{per\,cent}}
\newcommand{\Gyr}{\,\mathrm{Gyr}}
\newcommand{\kpc}{\,\mathrm{kpc}}
\newcommand{\dif}{\mathop{}\!\mathrm{d}}

\newcommand{\dddr}{\ensuremath{\dif^{3}\!\boldsymbol{r}\:\:}}

\newcommand{\Beta}{\mathcal{B}}

\newcommand{\rs}{r_\mathrm{s}}
\newcommand{\riso}{r_\mathrm{iso}}
\newcommand{\nmax}{n_\mathrm{max}}
\newcommand{\lmax}{l_\mathrm{max}}
\newcommand{\rt}{r_\mathrm{t}}
\newcommand{\rvir}{r_\mathrm{vir}}

\newcommand{\A}{`A'\ }
\newcommand{\B}{`B'\ }

\newcommand{\wyn}[1]{{\textcolor{black}{#1}}}
\newcommand{\nwyn}[1]{{\textcolor{black}{#1}}}
\newcommand{\ev}[1]{\textcolor{black}{#1}}
\newcommand{\jason}[1]{\textcolor{black}{#1}}

\title[Distorted and Evolving Haloes]{Models of Distorted and Evolving Dark Matter Haloes}

\author[Sanders, Lilley, Vasiliev, Evans \& Erkal]{
Jason L. Sanders,$^{1,2}$\thanks{E-mail: jason.sanders@ucl.ac.uk; ejl44,vasiliev,nwe@ast.cam.ac.uk}, 
Edward J. Lilley,$^{1}$,
Eugene Vasiliev$^{1,3}$,
N. Wyn Evans$^{1}$, 
Denis Erkal$^{4}$
\\
$^{1}$Institute of Astronomy, University of Cambridge, Madingley Road, Cambridge CB3 0HA, UK\\ 
$^{2}$Department of Physics and Astronomy, University College London, London WC1E 6BT, UK\\
$^{3}$Lebedev Physical Institute, Leninsky prospekt 53, Moscow, 119991, Russia\\
$^{4}$Department of Physics, Faculty of Engineering and Physical Sciences, University of Surrey, Guildford GU2 7XH, UK
}

\date{Accepted XXX. Received YYY; in original form ZZZ}

\pubyear{2020}

\begin{document}
\label{firstpage}
\pagerange{\pageref{firstpage}--\pageref{lastpage}}
\maketitle

\begin{abstract}
We investigate the ability of basis function expansions to reproduce the evolution of a Milky Way-like dark matter halo, extracted from a cosmological zoom-in simulation. For each snapshot, the density of the halo is reduced to a basis function expansion, with interpolation used to recreate the evolution between snapshots. The angular variation of the halo density is described by spherical harmonics, and the radial variation either by biorthonormal basis functions adapted to handle truncated haloes or by splines. High fidelity orbit reconstructions are attainable using either method with similar computational expense. We quantify how the error in the reconstructed orbits varies with expansion order and snapshot spacing.
Despite the many possible biorthonormal expansions, it is hard to beat a conventional Hernquist-Ostriker expansion with a moderate number of terms ($\gtrsim15$ radial and $\gtrsim6$ angular). As two applications of the developed machinery, we assess the impact of the time-dependence of the potential on (i) the orbits of Milky Way 
satellites, and (ii) planes of satellites as observed in the Milky Way and other nearby galaxies. Time evolution over the last 5 Gyr introduces an uncertainty in the Milky Way satellites' orbital parameters of $\sim 15 \,\percent$, comparable to that induced by the observational errors or the uncertainty in the present-day Milky Way potential. On average, planes of satellites grow at similar rates in evolving and time-independent potentials.
There can be more, or less, growth in the plane's thickness, if the plane becomes less, or more, aligned with the major or minor axis of the evolving halo.
\end{abstract}


\section{Introduction}

Cosmological dark matter haloes have a density law that has an approximate double-power form. This was first suggested by \citet{Du91}, but made famous by \citet{NFW1997}, who introduced the eponymous NFW density profile,
\begin{equation}\label{eq:nfw}
    \rho(r) = \frac{\rho_0 \rs^3}{ r(r+\rs)^2},
\end{equation}
where $\rs$ is the scalelength and $\rho_0$ the density normalisation \citep[see e.g.,][for a useful summary]{MBW}. Subsequent work showed that the slopes of the inner and outer power laws have some scatter about the canonical NFW values~\citep{Mo98,Kl01,Di2014,Dekel2016}, and even that the logarithmic gradient of the density slope may change with radius leading to an Einasto profile rather than double-power laws~\citep{Ei89,Merritt06}.

Even so, these laws are really no more than convenient fitting formulae that provide a zeroth order approximation to the dark halo density. Cosmological simulations have long shown that dark haloes are more complicated than simple spherical models. Triaxiality, shape or ellipticity variations with radius, substructure and lop-sidedness are all manifestations of the hierarchical assembly of galaxies via merging and accretion~\citep[e.g.,][]{Mo99,Jing2002,Pr19}. This has detectable consequences -- for example, streams caused by dwarf galaxies and globular clusters disrupting in lumpy haloes have markedly different morphologies to those disrupting in smooth haloes with idealized profiles~\citep[e.g.][]{Ngan2015}. 
Observationally, too, there are now clear indications that dark haloes have rich and complex shapes, which encode the physical processes that made them.  The modelling of long thin streams in the Milky Way halo such as the Orphan Stream has shown the importance of the gravitational effects of the Large Magellanic Cloud~\citep{Erkal2018}. This large satellite galaxy is in the process of merging with the Milky Way, and its gravitational pull causes both tidal distortions in the halo and reflex motion of the halo centre.  Equally, the stream from the disrupting Sagittarius galaxy in the Milky Way cannot be fit by a potential with fixed triaxial shape~\citep[c.f.,][]{La10,Be14}, but requires the dark halo shape to change from oblate to triaxial in the outer parts~\citep{Ve13}. \wyn{Similarly, the absence of fanning in the Palomar 5 stream suggests an almost axisymmetric potential in the inner Milky Way, whereas strong triaxiality is required to reproduce the morphology of the Sagittarius stream~\citep{Pe15}.} The description of the kinematics of stars and dark matter in the Milky Way galaxy then requires a much more elaborate dark matter potential than just a static, symmetric halo model with fixed shape.

There are also abundant applications to extra-galactic astronomy. For example, the reconstruction of low surface brightness features around nearby galaxies, the modelling of haloes of strong lenses at medium redshift and the kinematics of tracers like globular clusters and planetary nebulae around giant ellipticals all necessitate accurate representations of dark matter haloes. Deep multiband photometry from the {\it Vera Rubin Observatory} (first light in 2022) will open up a new domain of low surface brightness studies of stellar haloes of galaxies across a wide range of distances, masses and types~\citep[e.g.,][]{Laine}. These will be more difficult to model compared to structures in the Milky Way Galaxy, as the data are limited to two dimensions, but much can still be learned about the properties of dark haloes from ensemble modelling.

This paper develops the idea of describing dark matter haloes using basis function expansions. The potential and density are written as
\begin{equation}
\begin{split}
\Phi(\boldsymbol{r}) &=  \sum_{lm} \: B_{lm}(r) Y_{lm}(\theta,\phi), \\
\rho(\boldsymbol{r}) &=  \sum_{lm} \: A_{lm}(r) Y_{lm}(\theta,\phi),
\end{split}
\label{eq:crux}
\end{equation}
where the $Y_{lm} (\theta,\phi)$ are the unit-normalized spherical harmonics. Two approaches for representing the radial dependence of the spherical-harmonic coefficients, $A_{lm}(r)$ and $B_{lm}(r)$, have been explored in the literature: either as a weighted sum of orthonormal functions involving polynomials of degree $n$ in a scaled radial variable, or as interpolated functions defined by values at $n$ radial grid points. By far the most well-known of the former methods is the \citet{hernquist1992} expansion, although the method has its genesis in the earlier work of \cite{cluttonbrock1972}. The latter method can be traced back to N-body integrators \citep[e,g.,][]{Aarseth67}, but its implementation in this context has been advocated recently by \cite{Vasiliev2013}. In principle, these expansions can encode complex shape variation, together with arbitrary inner and outer density fall-offs of the halo. They therefore can represent elaborate 
time-independent potentials. However, the methodology is capable of still greater flexibility if the spherical-harmonic coefficients are also made functions of time, $A_{lm}(r,t)$ and $B_{lm}(r,t)$. This allows the representation of distorted and evolving dark matter haloes, which may be affected by time-dependent perturbations such as tidal interactions with other nearby galaxies.

The variety of applications for any basis function method is very substantial, as already articulated clearly by \citet{lowing2011} and \citet{Ngan2015}. If different snapshots of a numerical simulation are expressed in basis function expansions, the time evolution of the simulation can be recreated using interpolation. The simulations can then be replayed speedily many times with completely new (low-mass) objects inserted. This makes it ideal for studying myriads of problems in galaxy evolution and near-field cosmology, including the disruption of satellites and subhaloes, the precession of tidal streams or planes of satellites, and the build-up of the stellar halo. Provided the mass ratio of accreted object to host halo is less than 0.1, the effects of dynamical friction are unimportant~\citep{Bo08} and the inserted object has no back reaction on the rest of the simulation. 
However, to convert this powerful idea into an efficient working tool requires addressing a number of questions:

\begin{enumerate}
\item
Cosmological haloes participate in the large-scale Hubble flow. They are not isolated but feel the external tidal forces from larger scale structure, as well as the buffeting of frequent accretion events. The integration of orbits in the basis function expansion must also take account of these effects, if the orbits in the simulation are to be recovered accurately. How should they be modelled?

\item 
Suppose snapshots of an evolving numerical halo are available at fixed times as basis function expansions. An approximation to the state of the halo at intermediate times is recovered by interpolating the coefficient of each basis function between the preceding and following snapshots~\citep{lowing2011}. What is the best choice of time interval between snapshots and interpolation scheme? This can be answered by comparing orbits integrated in this time-varying basis function approximation with the original $N$-body trajectories. 

\item
Which expansion is optimal for a given simulation? Previous applications of this idea have routinely used the familiar Hernquist-Ostriker biorthonormal expansion \citep[e.g.,][]{lowing2011,Ngan2015}, but there are now many more options available~\citep[e.g.,][]{Vasiliev2013,LSE,Lilley2018b}. This necessitates the development of an error measure for the evolving haloes, based on the fidelity of orbit reconstruction to assess the competing methods.
\end{enumerate}

This paper provides answers to all these questions. It is arranged as follows. Section~\ref{sec:recap} recapitulates the biorthonormal and spline expansion methods. Section~\ref{sec:build} explains in detail the construction of both basis function expansions for one numerical halo, describing their usage in a time-evolving setting. Section~\ref{sec:perf} discusses the accuracy of the resulting halo representations, and which parameters can be adjusted in order to achieve the optimal speed/accuracy trade-off. Finally, in Section~\ref{sec:app} we discuss some applications of the evolving halo model, including modelling the orbits of Milky Way satellites and the longevity of planes of satellites in Milky-Way-like haloes.


\section{Basis Function Methods}
\label{sec:recap}

There are two choices for the radial dependence of the spherical-harmonic coefficients, $A_{lm}(r)$ and $B_{lm}(r)$, introduced in equation~\eqref{eq:crux}. Both methods express the coefficients as convergent series indexed by $n$ that is truncated at order $n=n_\mathrm{max}$. The first represents each term as a weighted sum of biorthonormal functions of degree $n$ expressed in terms of a scaled radial variable. The second method uses interpolating functions defined by values at an arbitrary set of $n$ radial grid points. Both expansions can encode complex shape variation, together with arbitrary inner and outer density fall-offs of the halo, and become increasingly accurate with increasing $n$. 

\subsection{Biorthonormal Expansions}

In the first approach, using an biorthonormal basis function expansion or basis set expansion (BSE), we write equation~\eqref{eq:crux} as
%
\begin{align}\label{eqn:bfe_series}
\Phi(\boldsymbol{r}) &=\sum_{nlm} \: C_{nlm} \: \Phi_{nlm}(\bs{r})=  \sum_{nlm} \: C_{nlm} \: \Phi_{nl}(r) \: Y_{lm}(\theta,\phi),\\
\rho(\boldsymbol{r}) &=\sum_{nlm} \: C_{nlm} \: \rho_{nlm}(\bs{r})= \sum_{nlm} \: C_{nlm} \: K_{nl} \: \rho_{nl}(r) \: Y_{lm}(\theta,\phi).
\end{align}
%
The basis functions are normalized to ensure
Poisson's equation is satisfied as
\begin{equation}
    \nabla^2\Phi_{nlm}(\bs{r}) = 4\pi G\rho_{nlm}(\bs{r}).
\end{equation}
The biorthonormality condition implies that
\begin{equation}
\int \dddr \Phi_{nlm}(\boldsymbol{r}) \: \rho_{n^\prime l^\prime m^\prime}(\boldsymbol{r}) = K_{nl} \: N_{nl} \: \delta_{nlm}^{n^\prime l^\prime m^\prime}.
\end{equation}
This method is efficient if the expansion captures at zeroth-order the spherically-averaged density profile of a cosmological dark-matter halo. Any deviations are then succinctly described by a small number of the higher-order terms in the basis set. This idea was introduced in \citet{cluttonbrock1972,cluttonbrock1973}. The most well-known example is the \cite{hernquist1992} expansion (hereafter HO), which uses the simple \citet{Hernquist1990} model as its zeroth order term.  Subsequently, \citet{zhao1996} found a neat way to generalise this to expansions with the hypervirial models~\citep{Ev06} as the lowest order term. \wyn{\citet{weinberg1999} presented a numerical algorithm that constructed sets of biorthonormal basis functions that followed any
underlying spherical profile. Matters then laid in abeyance until recent work~\citep{LSE,Lilley2018b} considerably extended the classes of models with analytic biorthonormal basis function expansions.} These new expansions give greater freedom in the choice of zeroth-order inner and outer slope, and completely encompass all previously discovered spherical expansions as special cases. The apparatus now exists to use more general double power-law density models as the zeroth order, about which we can build expansions describing a rich variety of shapes and profiles. 

Biorthonormal expansions have some considerable advantages: 
\begin{inparaenum}
    \item The recurrence relations for orthogonal polynomials used in the expansion enable the higher order basis functions to be calculated rapidly from the low order ones,
    \item The reconstructed potential and density are represented by the same coefficients, which are easy to compute as weighted integrals over the target density. When the density is formed from a cloud of point particles of mass $m_i$ as in a numerical simulation, the integral for the coefficients reduces to a sum over particles~\citep[e.g.,][]{LSE}
\begin{equation}\label{eq:particle_sum}
C_{nlm} \propto \int \dddr \: \Phi_{nlm}(\boldsymbol{r})\:\rho(\boldsymbol{r}) = \sum_i m_i\:\Phi_{nlm}(\boldsymbol{r_i}).
\end{equation}
The biorthonormality ensures that all the calculations are linear with respect to the particles.  
\item Because the potential, forces and density are all linear with respect to the same set of coefficients, we may compute the force at any intermediate moment of time by interpolating the coefficients in a suitable way, without increasing the cost of potential evaluation.


\end{inparaenum}

\subsection{Spline Expansions}

The second approach is to represent the radial dependence of each term in the expansion explicitly as interpolating functions on a radial grid. This idea has its roots in $N$-body simulations (e.g., \citealt{Aarseth67}, \citealt{McGlynn84}, \citealt{Sellwood03}, \citealt{Meiron14}), and as a computationally inexpensive way of solving the Poisson equation in Schwarzschild or made-to-measure modelling (\citealt{ValluriME04}, \citealt{deLorenzi07}, \citealt{Siopis09}), often with the restriction to axisymmetry. The coefficients of the angular spherical harmonic expansion are evaluated at a small set of radial grid points, and the radial dependence of forces is then interpolated (typically linearly) between grid nodes. 

\citet{Vasiliev2013} suggested using splines to represent the radial basis functions $A_{lm}(r)$ and $B_{lm}(r)$ in equation~\eqref{eq:crux}. In the most recent version of the algorithm \citep{Va19}, the potential coefficients are represented by quintic splines in a suitably scaled radial coordinate, so that the derivatives of the potential up to second order are twice continuously differentiable. The number and positions of nodal points can be chosen arbitrarily (typically a logarithmic radial grid is used), so the method is in principle very flexible. To construct a potential from a given smooth density profile, the latter is expanded in spherical harmonics, and then the Poisson equation is solved by 1d radial integration of each term. When an $N$-body snapshot is used as an input, the spherical-harmonic expansion of its density profile is constructed by penalized least-square fitting, as detailed in the appendix of~\citet{Va18}. Differently from a biorthonormal basis function expansion, the evaluation of the potential and forces at a given point depends only on the coefficients at a few nearby nodes rather than on the whole basis set. However, the computations are no longer linear because of the need for penalized least squares and due to various scaling transformations designed to improve the accuracy of interpolation.

\begin{figure*}
  \centering
  \subfigure{\includegraphics[width=.64\textwidth]{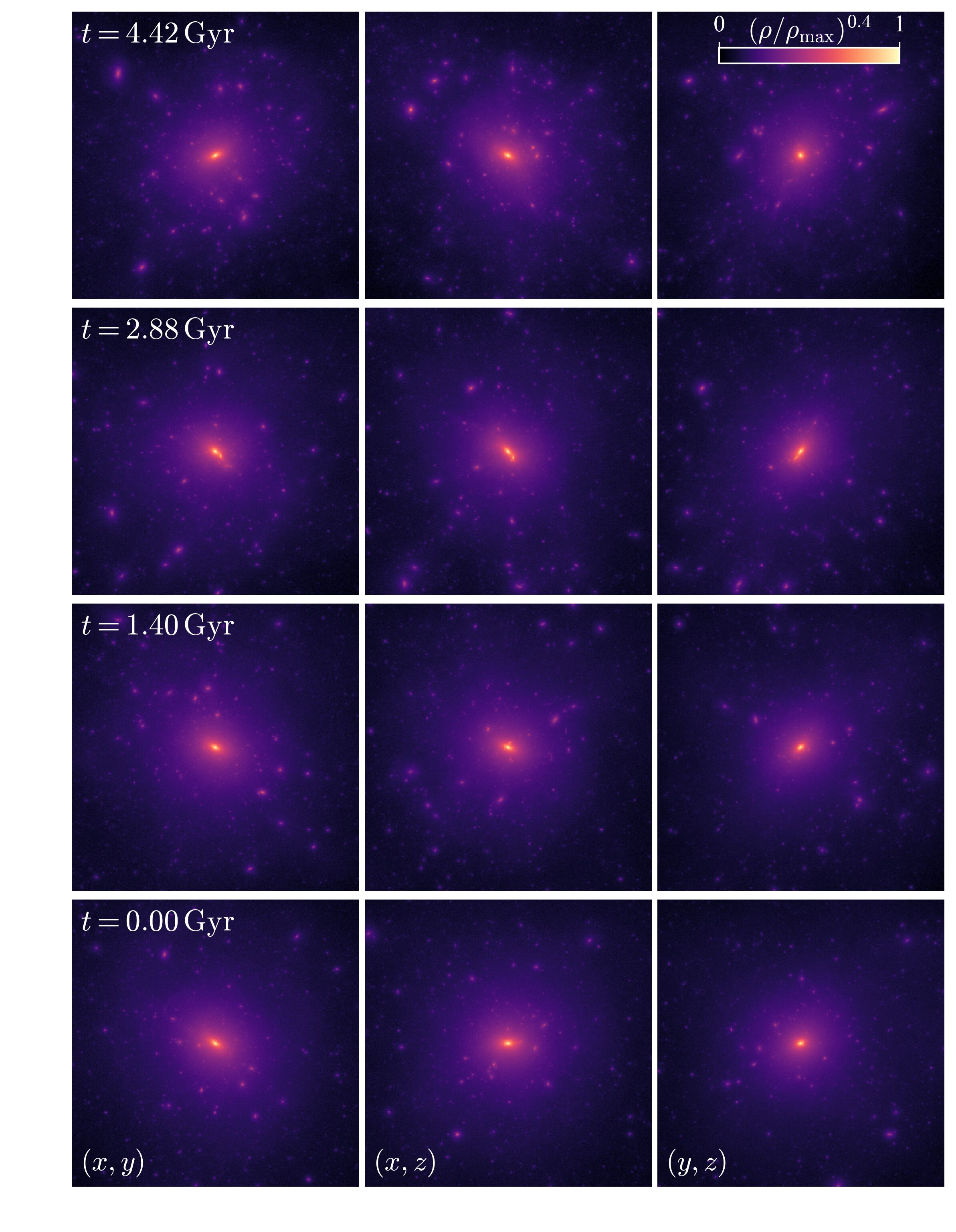}}
  \subfigure{\includegraphics[width=.35\textwidth]{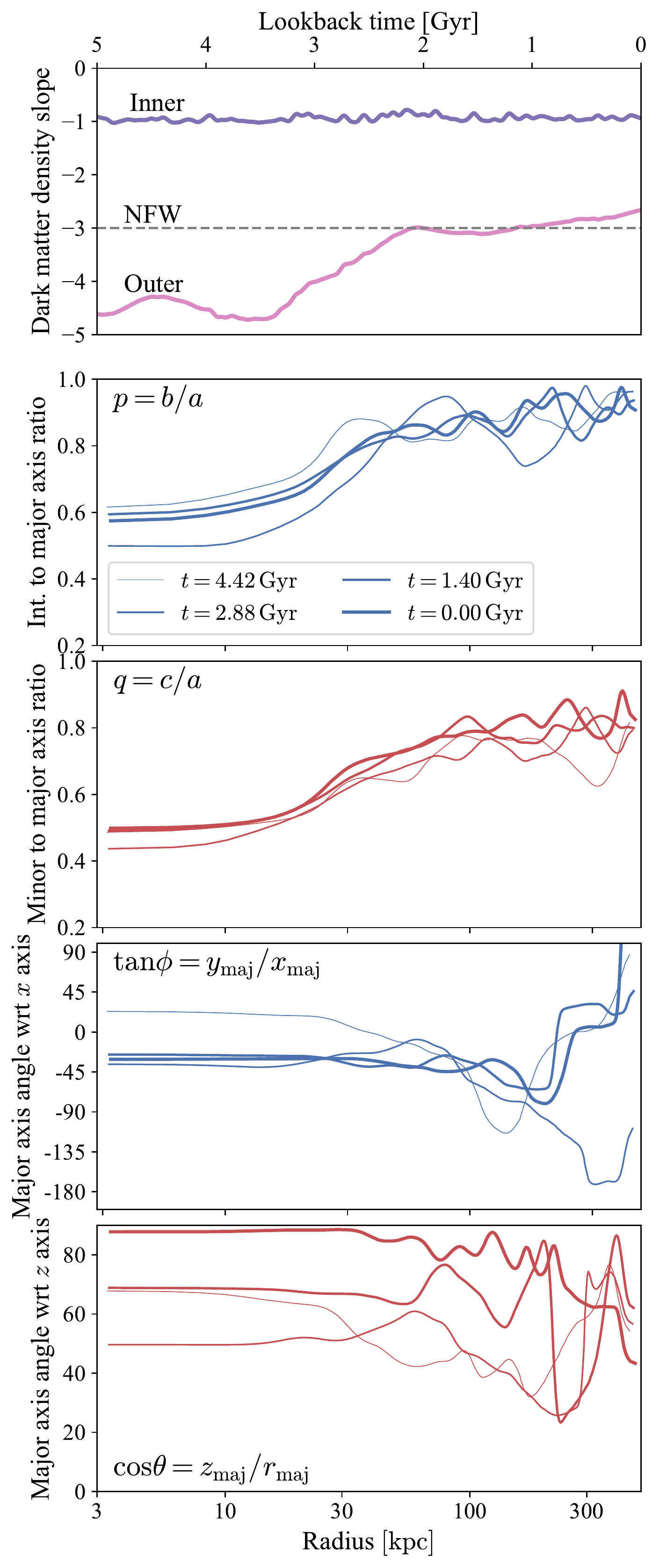}}
  \caption{Density of the studied dark matter halo at four snapshots. The left set of images shows the projected density. Each column displays a $500\,\mathrm{kpc}\times500\,\mathrm{kpc}$ projection of the halo (left: $(x,y)$, middle: $(x,z)$, right $(y,z)$). Each row is labelled by the lookback time. Note the time dependence of the large scale morphology. On the right we display the dark matter density slope as a function of time in the top panel and in the bottom four panels we show the axis ratios and the direction of the major axis at each radii for the four snapshots (thicker lines are later times).}
  \label{fig:halo_today}
\end{figure*}

\begin{figure*}
  \centering
  \includegraphics[width=0.99\textwidth]{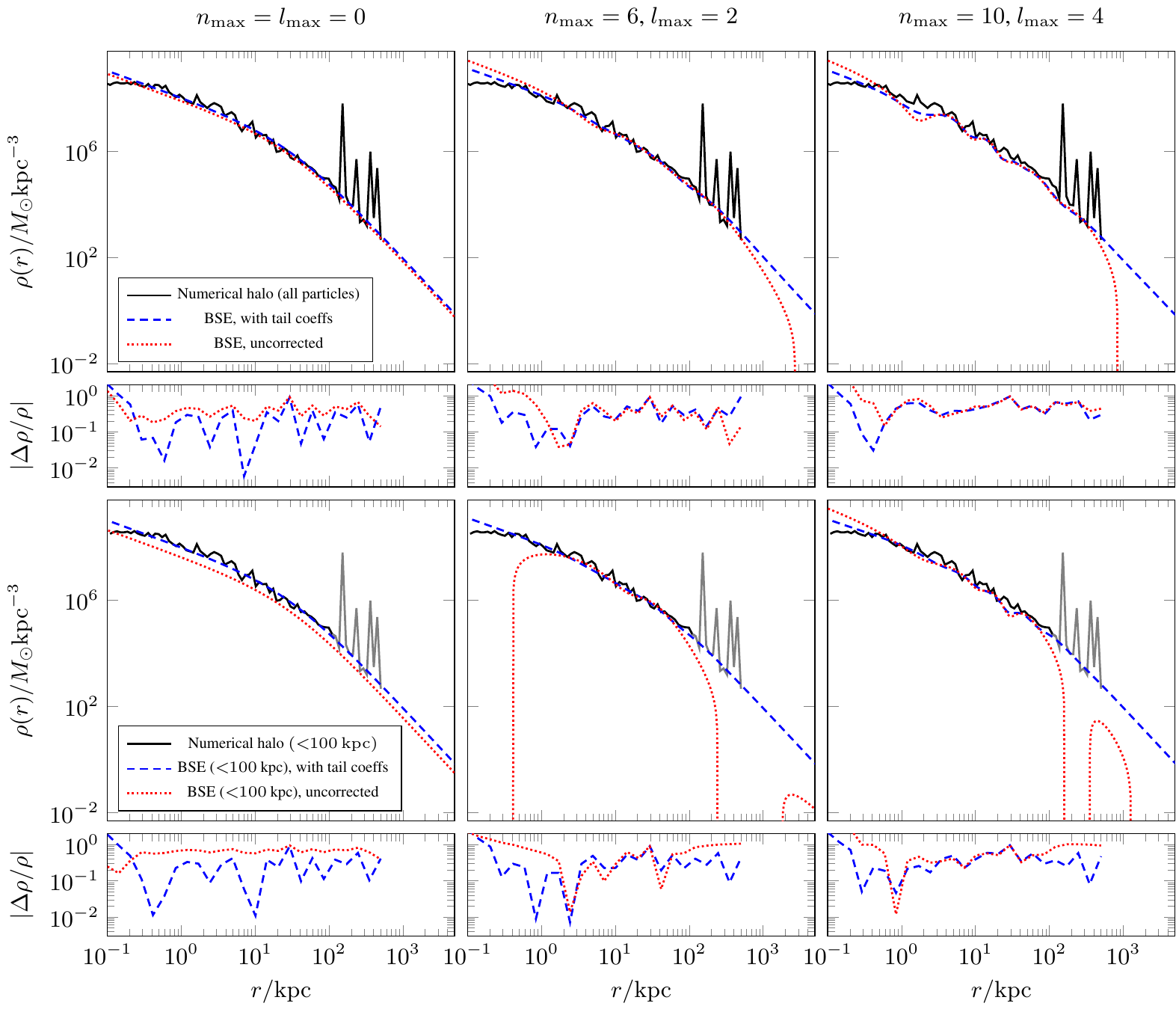}
  \caption{Radially-averaged density profiles of the halo at $11.8\Gyr$ (at which point it resembles an NFW profile), along with reconstructions using the BSE with parameters $\nu=0$, $\rs=30$ (the NFW case). The large spikes in the original halo density between 100-500 kpc correspond to substructure. The upper three panels show successively more detailed reconstructions, using $\nmax= \lmax =0$ (a single term), $\lmax=2$ (63 terms) and $\lmax=10$ (275 terms). The residuals are shown in the subpanels underneath. Notice that, without the tail correction, there are oscillatory artefacts in the density caused by the finite extent of the simulation data. The lower three panels magnify this effect by using approximately $1/4$ of the particles -- those from within $<100\kpc$ -- to compute the BSE coefficients. Neglecting the unresolved substructure, the corrected coefficients still provide a reasonable approximation outside of this range.}\label{fig:halo_profile}
\end{figure*}

\section{Application to a Time-evolving Halo}
\label{sec:build}

We now turn to the application of the two expansion methods to a simulated dark matter halo. We first describe the details of the simulation, before describing specific implementation choices for the two methods.

\subsection{A Milky Way-like dark matter halo}
\label{sec:halo_desc} 

\subsubsection{The Density of the Halo} 

The analysed simulations are run with the $N$-body part of \textsc{gadget-3}, which is similar to \textsc{gadget-2} described in \citet{springel_2005}. The zoom-in strategy follows \cite{onorbe} and all initial conditions are generated with \textsc{music} \citep{Ha11}. Cosmological parameters are taken from \cite{planck} with $h=0.679$, $\Omega_b = 0.0481$, $\Omega_0=0.306$, $\Omega_\Lambda=0.694$, $\sigma_8=0.827$, and $n_s=0.962$. In order to select a halo, we first simulate a 50$h^{-1}$ Mpc box with $512^3$ particles from $z=50$ to $z=0$. We use \textsc{rockstar} \citep{rockstar} to identify haloes and we select Milky Way-like haloes which have virial masses between $7.5\times10^{11}M_\odot-2\times10^{12} M_\odot$, no major mergers since $z=1$, and no haloes with half the mass of the Milky Way analogue's mass within $2h^{-1}$ Mpc. For a given halo, we select all particles within 10 virial radii and run an intermediate resolution zoom-in with maximum resolution $2048^3$, corresponding to a particle mass of $1.8\times10^6 M_\odot$. This intermediate step helps to reduce the contamination from low resolution particles in our final, high resolution zoom-in. For the final zoom-in, we take the intermediate resolution simulation and select all particles within 7.5 virial radii. We then run a zoom-in with a maximum resolution of $4096^3$, corresponding to a particle mass of $2.23\times10^5 M_\odot$. Our high resolution zoom-in is uncontaminated within $1h^{-1}$ Mpc of the main halo.

From the several simulated Milky Way-like haloes, we select a single halo as our benchmark model and we analyse it in detail in this paper. We focus on the final $5\,\mathrm{Gyr}$ of evolution as prior to this the evolution was more tumultuous and dominated by significant merger events. We record 67 evenly spaced snapshots for the initial $\sim8.8$ Gyr of the simulation and then we record snapshots every 10 Myr during the final 5 Gyr. The halo's density at four different snapshots since $5\,\mathrm{Gyr}$ ago is shown in Fig.~\ref{fig:halo_today}. The halo contains $1.3 \times 10^7$ particles and has a virial scale-length $\rvir = 325\kpc$ and a concentration $c=9.6$. We measure the inner and outer density slopes from a histogram of particles between $0.08$ and $4\,\mathrm{kpc}$, and $200$ and $500\,\mathrm{kpc}$ respectively. As shown in Fig.~\ref{fig:halo_today}, the inner slope is cusped with inner density slope $\gamma \approx 1$, consistent with the NFW model of equation~\eqref{eq:nfw}, whilst the outer slope evolves from a steeper fall-off of $\beta\approx4.5$ to the NFW value of $\beta=3$ for the last $2\,\mathrm{Gyr}$ of evolution. 
At all times, the halo has an approximately triaxial density distribution characterised by axis ratios $p$ in the $(x,y)$ plane and $q$ in the $(x,z)$ plane. The semiaxes $p$, $q$ and the direction of the major axis as a function of radius at four snapshots are shown in Fig.~\ref{fig:halo_today}. These quantities are computed from the moment of inertia for particles binned by their local density (employing a local density threshold to remove subhaloes). At all times, the central parts of the halo are more flattened ($p\approx0.6$ and $q\approx0.5$) than the outer parts ($p\approx0.9$ and $q\approx0.8$). The shape evolution over time is quite mild but the alignment of the major axis shows the halo tumbles significantly over the last $\sim4\,\Gyr$.

\subsubsection{The Forces on Particles}
Any method that reconstructs the force on each particle in the halo must contend with the fact that the halo is a non-inertial reference frame, as the centre of the coordinate system is at each step centred according to the cusp of the density distribution as found using \textsc{rockstar}. We here detail the computation of the fictitious force arising from the non-inertial frame.

%
%
%
%
The comoving coordinate of the halo centre is $\boldsymbol{x}(t) \equiv \boldsymbol{r}(t) / a(t)$, where $a(t)$ is the cosmological scale factor. The peculiar velocity (the physical velocity $\dot{\boldsymbol{r}}$ minus the Hubble flow) is also reported by \textsc{rockstar}, but since it analyzes each snapshot independently, the reported velocities do not correspond to time derivatives of positions. We therefore compute the peculiar velocity of the halo centre as $\boldsymbol{u} = \dot{\boldsymbol{r}} - H(t)\,\boldsymbol{r} = \dot{\boldsymbol{x}}\,a(t)$, where $H(t)\equiv \dot a(t)/a(t)$ is the Hubble parameter. The acceleration of the reference frame associated with the halo centre is simply $\dot{\boldsymbol{u}}$, which is calculated numerically. The force on the test particle is therefore
\begin{equation}   \label{eq:force}
\boldsymbol{F}(\boldsymbol{x},t) = -\boldsymbol{\nabla}\Phi(\boldsymbol{x},t) - \dot{\boldsymbol{u}}(t).
\end{equation}
where $\Phi(\boldsymbol{x},t)$ is the halo potential (as reconstructed by a basis function expansion). Eq~(\ref{eq:force}) now accounts for the forces on the halo overall, including those due to large-scale structure in the cosmological simulation. It neglects tidal effects at the scale of the halo itself, since the corresponding term $\boldsymbol{x}(t)\,\ddot a(t)/a(t)$ is several orders of magnitude smaller than the total force. We stress the importance of taking the acceleration due to the non-inertial reference frame into account: without the second term in the above equation, the agreement between the trajectories computed in the smooth halo potential and the original $N$-body simulation is much worse.

We also noticed that the position of the halo centre reported by \textsc{rockstar} fluctuates on a short timescale, and the corresponding noise in the acceleration is dramatically amplified by taking the second derivative. We found it advantageous to construct a smooth approximation for the trajectory of the halo centre-of-mass (with a timescale for variation of order few hundred Myr), and use it to derive the acceleration. The mismatch between the halo centre-of-mass position and velocity reported by \textsc{rockstar} and derived from our smooth approximation needs to be taken into account when using particle coordinates to initialize the potential, and when comparing trajectories of test particles to those taken from the original simulation. \ev{Without this additional smoothing of the centre-of-mass trajectory, the reconstructed orbits deviated more strongly (up to a factor of two larger errors) from the original ones in the central parts, where the orbital timescales are short, but the outer parts were less affected.}

\subsection{Implementation: Biorthonormal Expansions}

\subsubsection{Choice of Expansion}
\label{sec:implementbe}

The biorthonormal expansions have a double power law density form for the zeroth order basis function $\rho_{000}$, namely
\begin{equation}
\rho_{000}(r)\propto \frac{1}{r^{\gamma}(\rs^{1/\alpha}+r^{1/\alpha})^{(\beta-\gamma)\alpha}},
\label{eq:doublepowlaw}
\end{equation}
where the three parameters $(\alpha,\beta,\gamma)$ describe the turn-over, outer slope and inner slope.  
The familiar NFW model corresponds to $(\alpha,\beta,\gamma) = (1,3,1)$.
\citet{Lilley2018b} showed that there are two families (\A and \B) of biorthonormal basis functions, lying on distinct, intersecting surfaces in the $(\alpha,\beta,\gamma)$ space. The models have two parameters, $\alpha$ and $\nu$. 
Here, $\alpha$ corresponds exactly to the double power law $\alpha$ parameter, whilst $\beta$ and $\gamma$ are related to $\nu$ via $\gamma=2-1/\alpha$ and $\beta=3+\nu/\alpha$ for Family `A' and  $\gamma=2-\nu/\alpha$ and $\beta=3+1/\alpha$ for `B'. 

\wyn{The one-parameter family of \citet{zhao1996} arises as the intersection of the \A and \B families. It is obtained from either family by setting $\nu = 1$ and leaving $\alpha$ arbitrary. This basis set lies along the ray $(\alpha,\beta,\gamma) = (\alpha,3+1/\alpha,2-1/\alpha)$ in the double power law parameter space. The case $\alpha =1$ corresponds to the \citet{hernquist1992} expansion. The subset of the \A family obtained by setting $\alpha=1$ gives basis sets corresponding to the `generalised NFW'~\citep{Ev06} models, lying along $(\alpha,\beta,\gamma) = (1,3+\nu,1)$. Notably, this gives basis sets with flexible outer slopes, including the important NFW case ($\nu = 0$) and Hernquist-Ostriker ($\nu=1$) cases. Both the Zhao and generalized NFW sequences have the advantage of potentials expressible in terms of elementary functions, as opposed to special functions.}



\wyn{We expect a judicious choice of the basis-set expansion parameters ($\alpha, \nu, r_s$) will provide gains in efficiency and accuracy. Here, we motivate our choice:}

\begin{enumerate}

    \item The value of $\alpha$ controls not just the width of the turn-over region, but also the spacing of the zeroes of the polynomials used in the higher-order terms of the expansion. The argument of the polynomials is $r^{1/\alpha}/(1+r^{1/\alpha})$ giving rise to a spacing of the zeros of $\Delta\ln r\sim4\alpha/(n+1)$ (assuming for small $n$ zeros are near $r\sim1$). This heuristic argument shows that that to achieve optimum accuracy requires $\alpha \approx 1$. \jason{This agrees with our tests which demonstrate} an acceptable range of around $\alpha = 0.7$--$2$, outside of which the expansions become inefficient. In practice, this limits the flexibility of the \citet{zhao1996} expansions
    (though the widely-used \citet{hernquist1992} expansion does correspond to the choice $\alpha=1$ which obeys this constraint). 
    
    \item In our experiments on the reconstruction of this dark matter halo, we find that flexibility in the density of the outer slope $\beta$ is more desirable than in the inner slope $\gamma$, as our halo has a constant inner slope $\gamma \approx 1$ for the entire time interval considered. For this reason, we restrict our attention to the generalised NFW models, with their single free parameter $\nu$. Appendix~\ref{sec:nfw} summarizes the essential formulae for the generalized NFW expansions from \citet{Lilley2018b}, which we make use in the rest of the paper.

\item There is a final independent parameter, the scale-length $\rs$.  In our experiments, we find that the scale-length in the expansion $\rs$ must be set to a reasonable value, $\rs\approx(\nu+1)\riso$, where $\riso  = \rvir/c$ is the radius at which the logarithmic slope of the (spherically averaged) halo density attains the isothermal value of $-2$. The expansion becomes severely inaccurate if $\rs$ is less than a few per cent of $\rvir$, but otherwise the exact choice of $\rs$ is not important.
\end{enumerate}

To summarise, in the remainder of the paper, we use biorthonormal expansions with a generalized NFW model at lowest order. These have $(\alpha,\beta,\gamma) = (1,3+\nu,1)$. So, there is a single free-parameter $\nu$ remaining that controls the fall-off of the dark halo density at large radii.  We will set this by examining the fidelity of reconstructed orbits in Section ~\ref{sec:perf}.

\subsubsection{The outer tails of the expansion}
\label{sec:truncation}

In practice, a snapshot of a simulated halo has a truncation radius $\rt$, beyond which there are no particles. This is artificially introduced due to our cutout scheme (our halo data is truncated at $\rt = 500\kpc$). The na\"ive use of biorthonormal expansions on this data results in artefacts: spikes of negative density at very large and very small radii are produced at higher expansion orders ($\nmax > 10$), in a manner analogous to the Gibbs phenomenon that occurs when a finite number of terms in a Fourier series is used to resolve a jump discontinuity. There is also a severe under-estimate of the radial acceleration using the first few series coefficients ($\nmax < 5$). This arises as by construction the total mass of the expansion matches the simulation. When the profile is truncated, the mass is reduced so the best-fitting double-power law model underestimates the truth. Examples of these artefacts are visible in the `uncorrected' curves in the upper panels of Fig.~\ref{fig:halo_profile}. The lower three panels amplify this effect by using only the $3.7\times 10^6$ particles found within $<100\kpc$ to compute the coefficients. The lower panels also illustrate another pitfall, namely that the basis expansion tries to reproduce the hard cut-off at $\rt$, rather than the desired asymptotic power-law behaviour. 
 
Our strategy for solving this problem is the extrapolation of the $N$-body data beyond the truncation radius $\rt=500\,\mathrm{kpc}$, assuming it follows a power law. This is accomplished by adding to each coefficient a fixed quantity $T_{nlm}$ -- multiple evaluations of the series do not require any additional calculations, so this computational effort scales only with the number of terms in the truncated series. Denoting the \lq uncorrected' coefficients by $C^\mathrm{orig}_{nlm}$, the corrected coefficients are
\begin{equation}
  C_{nlm} = C^\mathrm{orig}_{nlm} + \mathcal{A} T_{nlm},
\end{equation}
where $\mathcal{A}$ is a normalisation constant that ensures that the mass interior to a chosen radius matches that of the $N$-body data $M_\mathrm{enc}^{\mathrm{N-body}}$. This quantity is given by
\begin{equation}
  \mathcal{A} = \frac{M_\mathrm{enc}^{\mathrm{N-body}}(\rt) - M_\mathrm{enc}^\mathrm{orig}(\rt)}{M_\mathrm{enc}^\mathrm{tail}(\rt)}.
\end{equation}
Expressions for the quantities $T_{nlm}$, $M_\mathrm{enc}^\mathrm{orig}$ and $M_\mathrm{enc}^\mathrm{tail}$ may be found in Appendix \ref{sec:tail_coefficients} (see equations~\ref{eqn:Menc_expansion}--\ref{eqn:define_chi_In_Qj}), along with an argument motivating the method. We show the results of applying this procedure to the halo in Fig.~\ref{fig:halo_profile}, noting how the outer tail of the expansion is more reasonably handled and at all radii the density error is reduced (particularly noticeable when only considering particles with $r<100\,\mathrm{kpc}$).

This ruse of extrapolating the asymptotic power-law behaviour of the density beyond the truncation radius allows for the use of infinite-extent basis functions on a finite region. Previously the only \jason{analytic} basis functions for use on a finite region were the spherical Bessel functions \citep{polyachenko1981}, these having the disadvantage that they do not resemble any simple halo or bulge profile. \jason{The method of \cite{weinberg1999} allows for construction of Sturm-Liouville numerical basis expansions of finite extent which neatly avoid these issues.}

\begin{figure*}
    \includegraphics{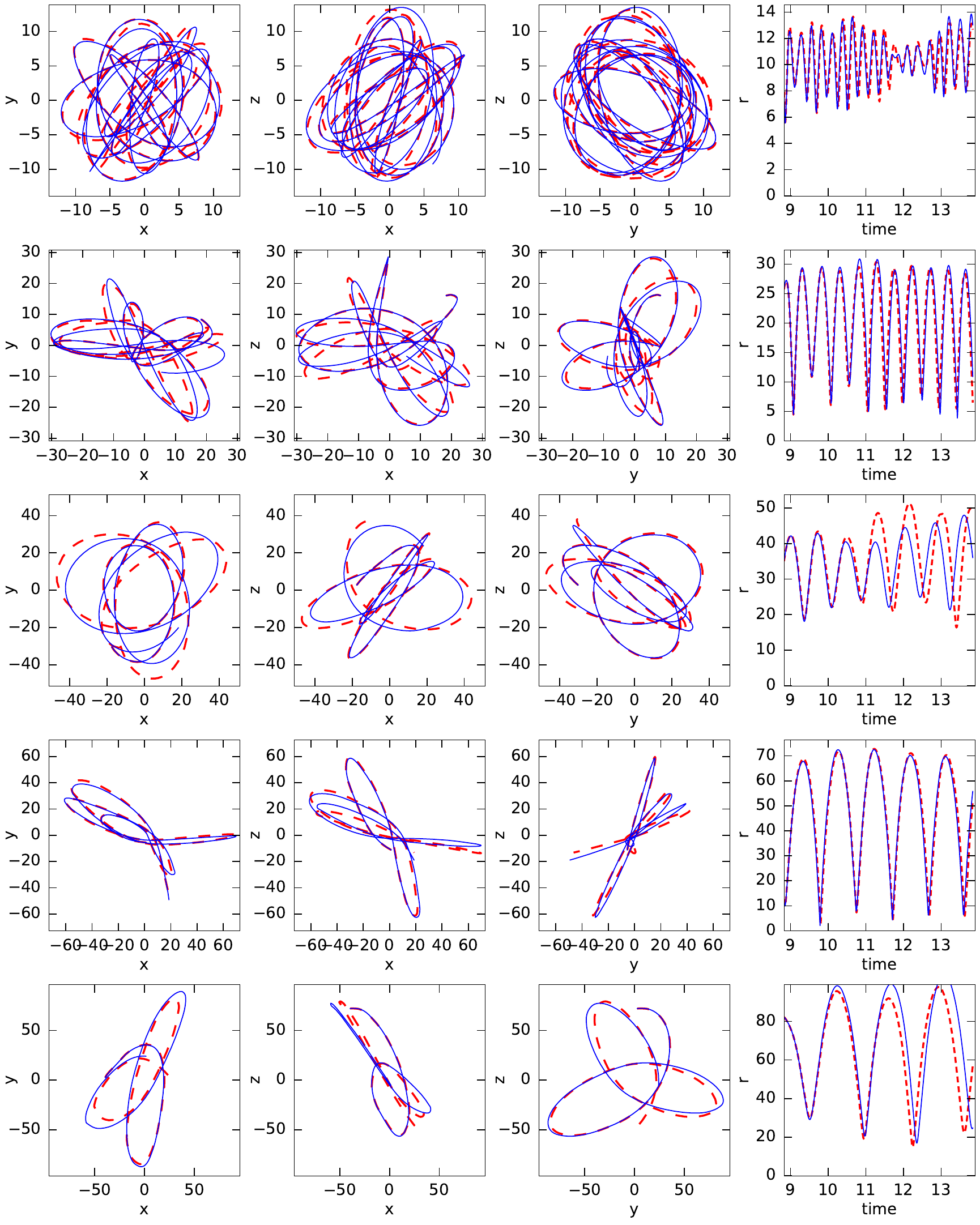}
    \caption{Examples of reconstructed orbits (blue) compared to the original trajectories of particles in the simulation (red dashed lines), for the spline method with $l_\mathrm{max}=10$. Each row plots a single orbit, with the first three columns showing its projections on three principal planes, and the last column -- time evolution of the galactocentric radius. Orbital period increases from top to bottom, and we illustrate both good cases (rows 1, 2 and 4), which are more common, and occasional bad reconstructions, usually caused by a single scattering event.}
    \label{fig:trajs}
\end{figure*}
\begin{figure*}
    \includegraphics[width=\textwidth]{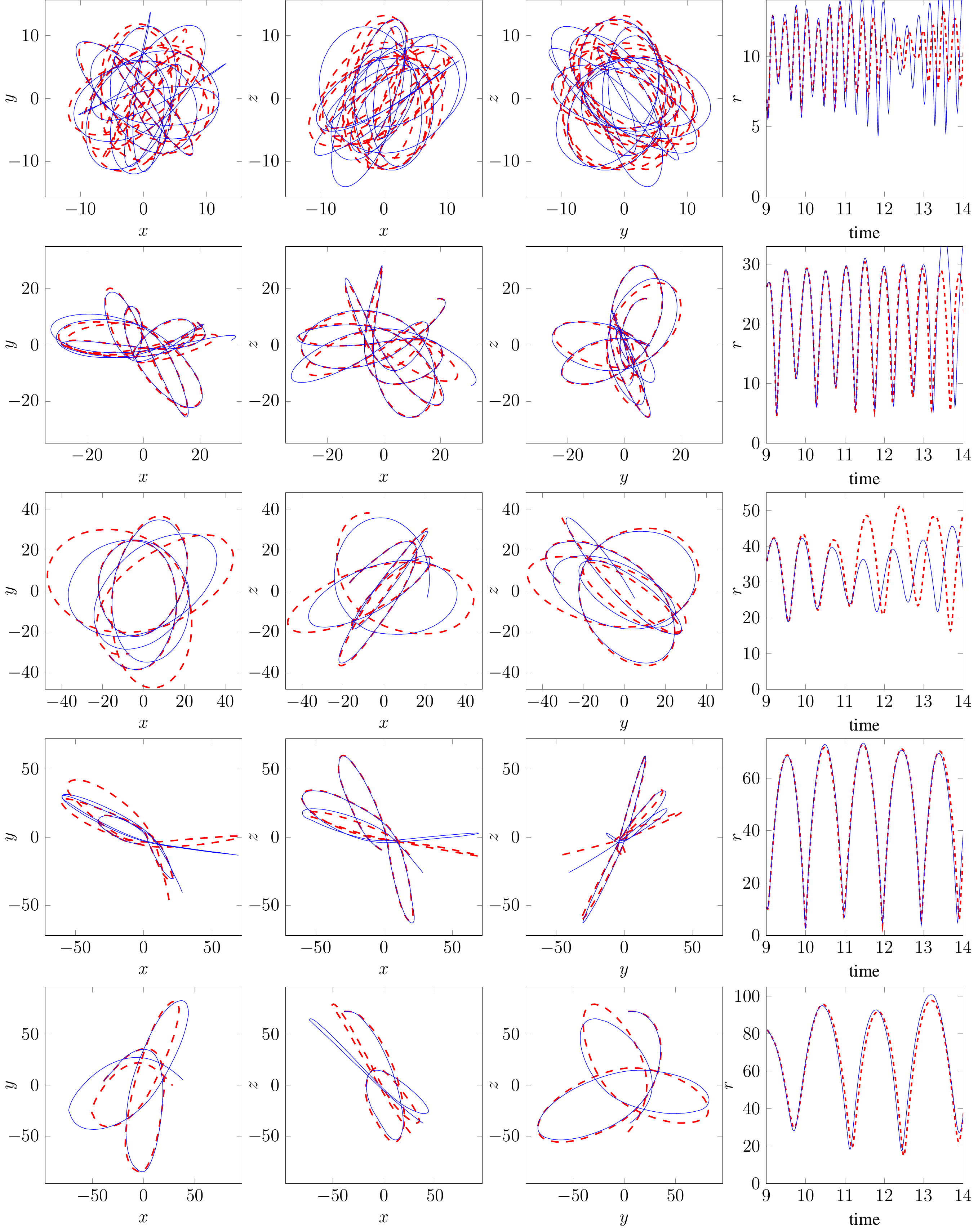}
    \caption{Similar to Fig.~\ref{fig:trajs}, but showing the results of the BSE reconstruction with $\lmax = 10$ and $\nmax = 22$ (blue) compared to the original trajectories (red dashes). Top to bottom show 5 different particles with increasing orbital period. Left to right show the three principal planes followed by galactocentric radius. As with the spline method there is a mix of good and bad reconstructions. The overall performance is very similar to the spline method.}
    \label{fig:trajs_bfe}
\end{figure*}
\begin{figure*}
  \centering
  \includegraphics[width=0.85\textwidth]{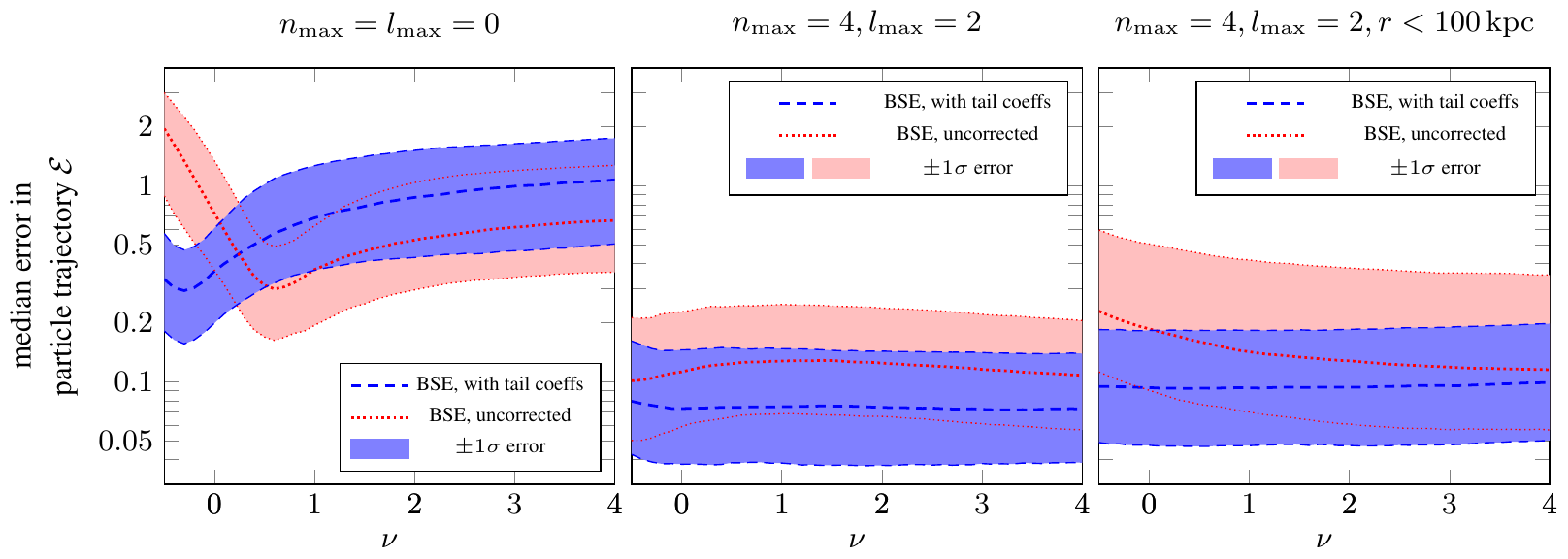}
  \caption{The run of the error measure $\mathcal{E}$ with the free parameter of the \jason{basis-set expansion} $\nu$. The outer logarithmic slope of the \jason{zeroth-order} density profile is $3+\nu$ such that an NFW halo has $\nu=0$. For each \jason{expansion} we fix the scale-length such that the isothermal length $\riso = 30\kpc$. The shaded region covers the 16th--84th percentile in $\mathcal{E}$. The dotted red \jason{(dashed blue)} lines refer to expansions which omit \jason{(include)} the `tail' correction discussed in Appendix~\ref{sec:tail_coefficients}. 
  \jason{The left panel shows results using only a single term in the expansion $(n_\mathrm{max}=l_\mathrm{max}=0)$, the middle panel uses $n_\mathrm{max}=4$, $l_\mathrm{max}=2$ corresponding to $45$ terms and the right panel uses the same number of terms but truncates the halo at $\rt = 100\kpc$.}
  For this figure we start the particle trajectories at $t\approx11.8\Gyr$ when the halo resembles an NFW halo.}
  \label{fig:choosing_parameters}
\end{figure*}

\subsection{Implementation: Spline expansions}

Unlike the biorthonormal expansion, in the spline-interpolated multipole approach implemented in \textsc{Agama} \citep{Va19}, the radial dependence of each spherical-harmonic term of the density expansion is represented by its values at a predefined grid of points in radius. There is still a considerable freedom in assigning the location of grid nodes, but the most natural choice is to use a uniformly-spaced grid in $\ln r$ with a fixed ratio between successive grid points $\mathcal R \equiv r_{i+1}/r_i$. In this case, the \lq relative resolution' (the radial extent of the smallest representable feature divided by its distance from origin) is constant across the entire system. In particular, the radial and angular resolutions roughly match when $\ln\mathcal R \approx 2.5/l_\mathrm{max}$. 
Typical grid sizes are $20-30$ radial points covering several decades in radius, and the accuracy starts to deteriorate remarkably when using less than 15 points.

To construct the smooth density profile from an $N$-body snapshot, \textsc{Agama} uses penalized spline fits with automatic choice of smoothing parameters. 
These procedures are detailed in the appendix of \citet{Va18}, and their cost is linear in both the number of particles and the size of the grid.

After a smooth multipole representation $A_{lm}(r)$ of the density is constructed, the corresponding potential terms $B_{lm}(r)$ and their radial derivatives at each grid node $r_i$ are computed by 1d integration:
\begin{equation}
\begin{array}{l} \displaystyle
B_{lm}(r_i) = \frac{4\pi\,G}{2l+1} \;\times \\[3mm]
\displaystyle \left[
r_i^{-l-1} \int_0^{r_i} A_{lm}(r)\, r^{l+2}\, dr + r_i^l \int_{r_i}^\infty A_{lm}(r)\, r^{1-l}\, dr
\right].
\end{array}
\end{equation}

The multipole terms of the potential are interpolated as 1d quintic splines in $\ln r$ defined by their values and derivatives at grid points. In doing so, the $l\ne 0$ terms are additionally scaled by the value of the $l=0$ term, and the latter is logarithmically scaled. Another, more efficient 2d quintic interpolation scheme is used when $l_\mathrm{max}>2$, representing each azimuthal Fourier harmonic term $B_m(r,\theta)$ on a 2d grid in $r,\theta$. All these scalings, together with the use of penalized spline fits for the density, break the linearity of the potential representation, but in practice the effect of this is negligible for a large enough $N$-body system.

The computational advantage of evaluating a spline-interpolated potential is that one needs to sum only $\mathcal O(l_\mathrm{max}\,m_\mathrm{max})$ or $\mathcal O(m_\mathrm{max})$ terms for 1d/2d interpolation schemes, respectively, instead of $\mathcal O(n_\mathrm{max}\,l_\mathrm{max}\,m_\mathrm{max})$ terms in the case of a radial basis set.
The cost of construction of a potential is broadly similar between the two approaches, since the need to evaluate rather expensive basis functions is comparable to the cost of spline fits. We will discuss this further in Section~\ref{sec:perf}.

\subsection{Time evolution}

Following the simulation over a range of times requires an approach to interpolating the potential expansions between the fitted snapshots. We adopt slightly different procedures for the two expansion methods.

In the biorthonormal expansion approach, we consider all the time-dependence in the gravitational force to be due to the series coefficients
\begin{equation}
\boldsymbol{F}_\mathrm{BE}(\boldsymbol{x},t) = -\sum_{nlm}C_{nlm}(t)\boldsymbol{\nabla}\Phi_{nlm}(\boldsymbol{x}),
\end{equation}
and so in order to get the force at intermediate times (say between halo snapshots at $t_1$ at $t_2$), we interpolate the coefficients,
\begin{equation}
C_{nlm}(t) = \tau(t)C_{nlm}(t_1) + \left(1 - \tau(t)\right)C_{nlm}(t_2),
\end{equation}
where $\tau(t)$ is a function that satisfies $\tau(t_1) = 1$ and $\tau(t_2) = 0$. For linear interpolation, we use
\begin{equation}
  \tau(t) = (t-t_2)/(t_1-t_2).
\end{equation}
This can be straightforwardly extended to higher-order interpolation such that the potential remains linear in the coefficients $C_{nlm}(t)$. For example, for cubic interpolation, the coefficients are evaluated at four consecutive times $(t_0,t_1,t_2,t_3)$. The function $\tau$ is then a Lagrange interpolating polynomial that depends on all four values $t_ik$.

As the acceleration is linear in the coefficients, the force due to interpolating the coefficients is equal to that which would result if we calculated the forces first and then interpolated. The fictitious force due to the halo reference frame $\dot{\bs{u}}$ is known in advance, and so is simply interpolated in the same way as the coefficients and added on at every time-step. The parameters of the expansion $(\rs,\nu)$ are chosen just once, using the first snapshot. 

In the spline approach, interpolation of coefficients between two snapshots is possible, but impractical due to additional costs associated with initializing the spline representation of the potential before computing forces. Instead, we may calculate the forces at the two nearest moments of time, and then linearly interpolate them to the intermediate time.
%
This scheme doubles the computational cost of potential evaluation in the time-dependent case. However, we find that in many cases a simpler approach of taking the potential from the nearest snapshot, without any interpolation, gives satisfactory results and does not increase the costs. The fictitious force corresponding to the non-inertial reference frame is interpolated in time as a cubic spline.

\section{Performance of our expansions: particle orbits}
\label{sec:perf}

With the implementation details established, we now turn to the question of how successfully the potential expansions can emulate properties of the simulation. In general, we want any expansion to successfully reproduce the paths of particles in the simulation, at least in a statistical sense. We therefore opt to inspect a fixed, but representative, sampling of particles within the simulation and test whether their orbits are reproduced \citep[see][for a similar discussion of the Aquarius simulations]{lowing2011}. We begin by defining our orbit sample.

\subsection{Orbits}

We consider a subset of particles from the original simulation satisfying the following criteria:
\begin{itemize}
\item the galactocentric radius never exceeds 200~kpc and is below 100~kpc in the last snapshot;
\item the orbital period is less than 3~Gyr;
\item the particle does not belong to any subhalo at the initial moment ($t \simeq 9$~Gyr), meaning that it is neither gravitationally bound to it nor resides within 10 scale radii of the subhalo.
\end{itemize}
Approximately $20\,\percent$ of all particles in the simulation satisfy these conditions, from which we randomly pick $\sim2000$ particles.

Figure~\ref{fig:trajs} shows five example orbits from the simulation, compared to the reconstructed orbits using the Spline expansion at the highest order \jason{inspected later} ($\nmax=40$, $\lmax=10$). The same orbits are shown in Fig.~\ref{fig:trajs_bfe} using the \jason{highest-order} basis-function expansion \jason{we consider} with $\nmax=22$, $\lmax=10$. We note in general the highly successful reproduction of the orbits over many orbital periods using both methods. For some orbits the spline method is superior (e.g. short-period orbits such as No.~1) whilst for others the basis expansion is better (e.g. long-period orbits such as No.~5). From visual inspection, the majority of orbits in our full sample are reproduced fairly well over many orbital periods, at least when considering overall orbit parameters such as the peri- and apocentre radii, although the actual trajectories start to diverge due to slight phase differences at later times. Occasionally, though, a particle from the original simulation may experience a close encounter with a subhalo or some other sudden perturbation not reproduced by the reconstructed orbit, after which these two trajectories diverge more strongly. Even though we illustrate these cases in two out of five panels, the actual occurrence rate of strong perturbations is more rare.

We now quantitatively inspect the reproduction of our chosen orbit sample, concentrating on  \begin{inparaenum}
    \item the difference between the two expansions,
    \item the variation in accuracy with specific parameter choices in the potential expansions,
    \item the accuracy with which different types of orbits are reproduced.
\end{inparaenum}
For this discussion, we require the introduction of a measure of the quality of orbit recovery.

\begin{figure*}
    \centering
    \includegraphics[width=.94\textwidth]{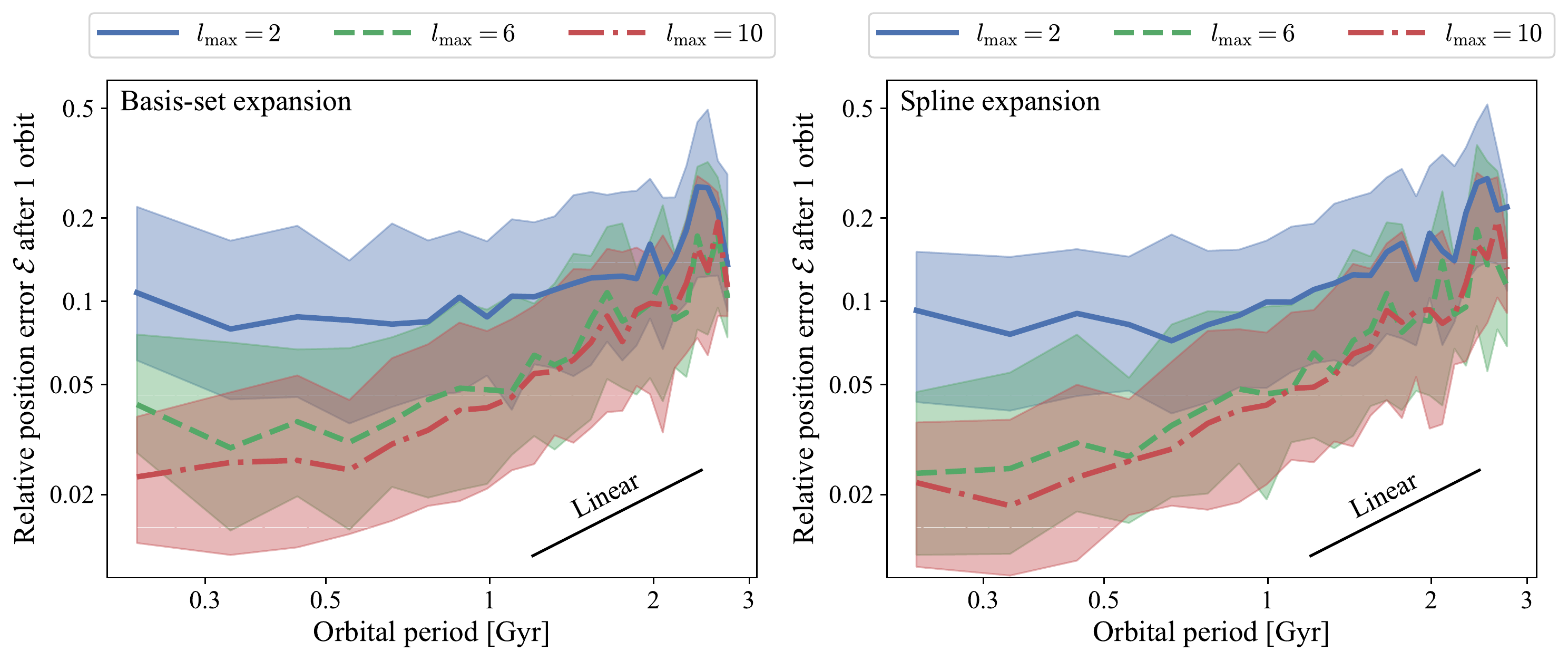}
    \includegraphics[width=.94\textwidth]{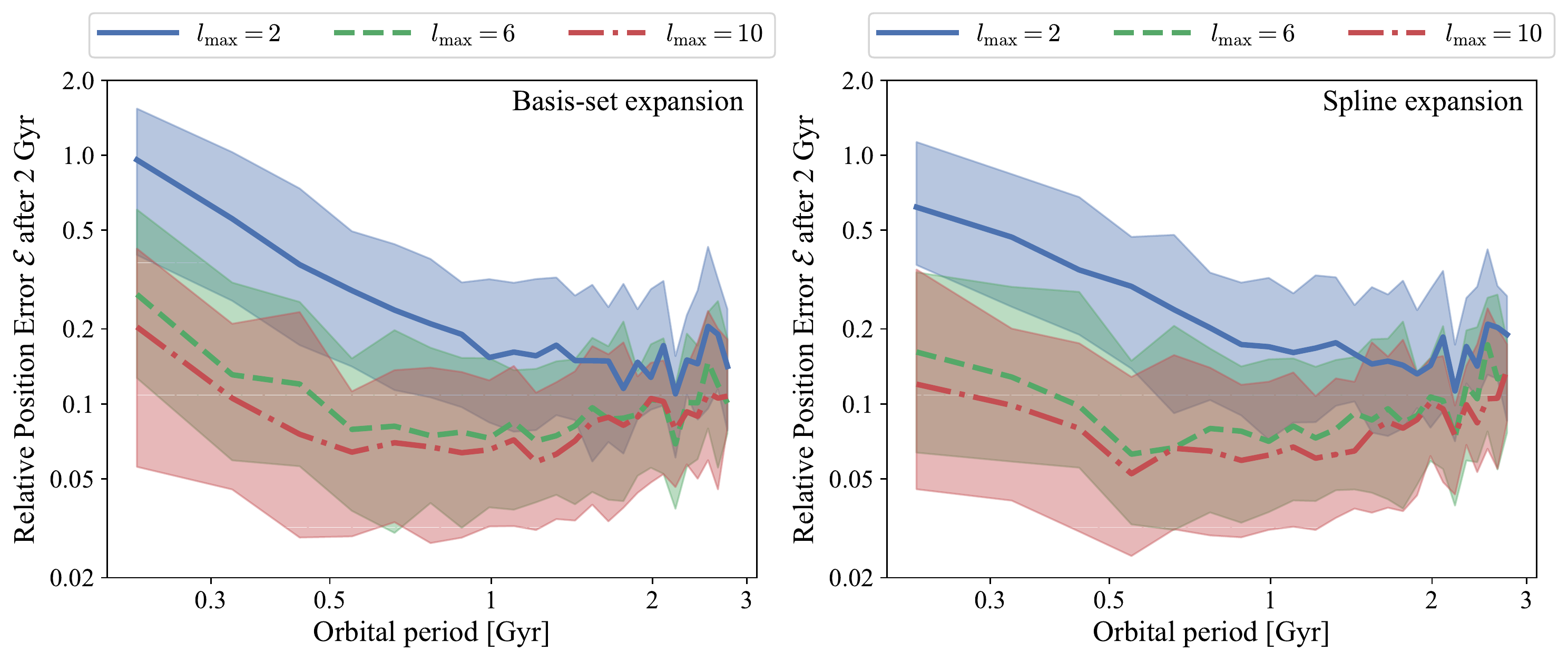}
    \caption{Median and $\pm1\sigma$ of relative position error after a single orbit (top) and after $2\,\mathrm{Gyr}$ (bottom) as a function of orbital period. The left panels show results for the basis function expansion and the right for the spline expansion. In each panel we show three sets of results: solid blue for $\lmax=2$, dashed green for $\lmax=6$ and dash-dotted red for $\lmax=10$. The corresponding number of radial terms ($n_\mathrm{max}$) is described in the text. The small black line shows a linear relation.}
    \label{fig:error}
\end{figure*}
\begin{figure*}
    \centering
    \includegraphics[width=\textwidth]{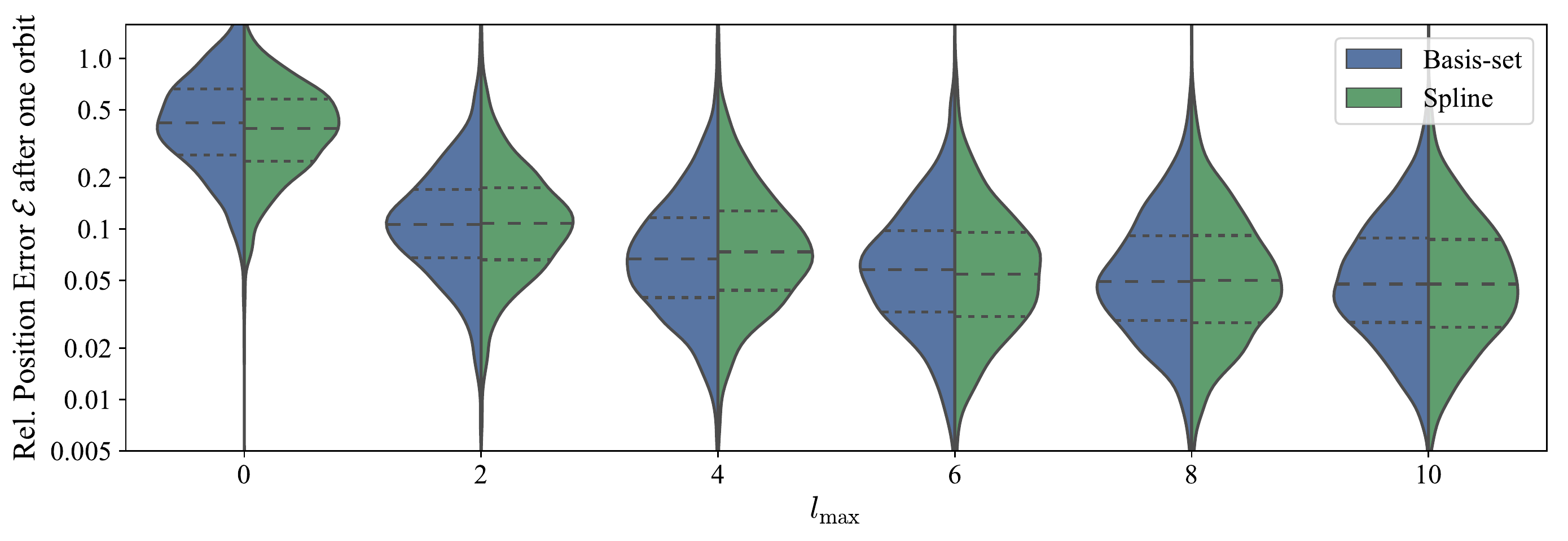}
    \caption{Violin plot showing the distributions of relative position error after one orbit. Left blue distributions are for the basis expansion method, and right green for the spline expansion method. The lines give the quartiles of the distributions (short-dashed are $25$th and $75$th and long-dashed $50$th).}
    \label{fig:violin}
\end{figure*}

\subsection{Error Measure}

The error in reconstructions has previously been studied with the mean integrated square error or MISE~\citep[e.g.,][]{Hall1983,Silverman,Vasiliev2013}. This involves integrating the squared magnitude of the absolute difference in the density or the acceleration field between the reconstructed and original halo over its entire spatial extent. This is best suited to static haloes rather than evolving ones. \ev{While it is clear that a lower error in acceleration generally corresponds to a better approximation, the relation between MISE and the accuracy of orbit reconstruction is not straightforward.}

Therefore, in order to test the fidelity of a given potential expansion of a time-evolving halo, we instead use the \emph{relative position error} of reconstructed orbits.
We define the relative position error of the reconstructed spatial path of the $i$th orbit $\vec{r}_{\mathrm{recon},i}(t)$ from the truth $\vec{r}_{\mathrm{orig},i}(t)$ after a time interval $t_i$ as
\begin{equation}\label{eqn:traj_error}
  \mathcal{E}_i =
\frac{\left\|\vec{r}_{\mathrm{orig},i}(t_i) - \vec{r}_{\mathrm{recon},i}(t_i)\right\|}
{r_{\mathrm{orig},i}}
.
\end{equation}
$r_{\mathrm{orig},i}$ is the time-averaged radius of the $i$th orbit.
We choose to perform the comparison after a single period $t_i=T_i$ for each orbit, although we will see this choice is somewhat arbitrary and using a fixed comparison time for all orbits produces qualitatively similar conclusions. 
$T_i$ is computed by taking a (zero-padded) fast Fourier Transform of the particle's original trajectory and computing one cycle with respect to the dominant frequency (if this time lies outside the simulated interval then time of the final snapshot is substituted). 
%

\subsubsection{Choice of Biorthonormal Expansion}

With a well-defined error measure selected, we are in a position to quantitatively select the optimal parameters for the biorthonormal basis expansion.
We have made preliminary choices already in Section~\ref{sec:implementbe}. Specifically, we argued that the expansion based on the \lq generalised NFW' models at zeroth order offered a good compromise between speed and realism. This is a one parameter family with $(\alpha,\beta,\gamma) = (1,3+\nu,1)$, so there remains a single parameter $\nu$ to be freely chosen.

The \wyn{left and middle} panels of Fig.~\ref{fig:choosing_parameters} show the median and $\pm1\sigma$ spread of $\mathcal{E}$ for our sample of orbits as a function of $\nu$. We show the results for the \lq tail corrected' (blue) and uncorrected (red) expansions, and consider evolution over the final $2\,\mathrm{Gyr}$ of the simulation. With just the zeroth order term ($\nmax=0$, $\lmax=0$), we expect the NFW model or $\nu=0$ to be preferred (see Fig.~\ref{fig:halo_today}) -- and such is the case for the corrected expansion. As the number of terms in the expansion increases to $\nmax=4$ and $\lmax =2$, the blue band becomes very flat, so the choice of $\nu$ is not at all important. There is no significant gain in using the expansion with the NFW model at zeroth order as compared to the simpler Hernquist-Ostriker expansion ($\nu=1$), for example. The main effect of the tail correction is to improve the median error, though there is a slight reduction in the width of the $1\sigma$ shaded region. 

\wyn{The right panel shows the effect of just using the particles in our sample that are within 100 kpc to construct the expansion. This exaggerates the importance of the artefacts in the uncorrected expansion, so we see larger discrepancies between blue and red bands. However, it is interesting that for the corrected coefficients with $\nmax=4$ and $\lmax =2$, there is little difference between the middle and right hand panels -- showing that we can use fewer particles ($3.7 \times 10^6$ of the particles are retained when truncating at 100 kpc, about a quarter of the total).} \nwyn{We have checked the dependence of  $\nmax$ and $\lmax$ on particle number, finding that the  accuracy does not deteriorate
even if particle number is reduced by a factor of a thousand.}

Based on these results, we use the Hernquist-Ostriker expansion ($\nu=1$) as our orthonormal basis series of choice, and proceed to examine its performance versus the spline expansion. 

\subsubsection{Comparison of the Biorthonormal and Spline methods}

In Fig.~\ref{fig:error}, we display the median and $\pm1\sigma$ spread of $\mathcal{E}$ for the considered sample of orbits as a function of their orbital period. For the two methods, we inspect the results for three choices of the number of angular terms: $\lmax=(2,6,10)$. \jason{From experiments with a wide range of $\nmax$, we found it prudent to set $\nmax=2\lmax+2$ as fixing $\lmax$ and increasing $\nmax$ (and vice versa) produced an $\nmax$ beyond which there was no improvement in accuracy. Our choice puts approximately equal resolution in angular and radial variations. For our three inspected cases, } this \jason{choice} corresponds to $n_\mathrm{max}=(6,14,20)$ radial terms for the basis-function expansion \jason{whilst for the spline expansion we more conservatively use} $(15,25,40)$ radial grid points (although the spline accuracy does not improve beyond 20 radial grid points, the computational cost is nearly independent of this number, so we allowed it to increase further). We note the increase in accuracy (reduction in $\mathcal{E}$) for increasing $\lmax$ particularly for the most bound orbits. We also observe that both methods perform similarly at equivalent order of expansion. This behaviour is further illustrated by Fig.~\ref{fig:violin} which shows the full distributions for $\mathcal{E}$ vs. $\lmax$. We see a rapid improvement in accuracy from $\lmax=0$ to $\lmax=4$ and a much slower improvement for higher $\lmax$. In general, the distributions of $\mathcal{E}$ are similar for the two methods and generically appear approximately Gaussian but with fatter tails particularly to high $\mathcal{E}$, probably due to particles scattered by subhaloes. 

The generic shape of the curves in Fig.~\ref{fig:error} (rising with increasing orbital period) is a result of our choice of time interval used in the evaluation of $\mathcal{E}$. Longer period orbits have their deviations measured over longer timescales so naturally accumulate more error. This is demonstrated by the approximate linear scaling of $\mathcal{E}$ with period. In Fig.~\ref{fig:error} we also display the distributions of $\mathcal{E}$ using a fixed time interval of $t_i=2\,\mathrm{Gyr}$. For this choice, we find the run of $\mathcal{E}$ is essentially flat for high periods and rises weakly at lower periods. However, the conclusions on the relative performance of different expansion orders are unchanged.

To summarize, we find that the accuracy of orbits improves with increase of the order of expansion, but only up to a certain limit ($\lmax \simeq 6$ and $n_\mathrm{max}\simeq 15-20$). We conjecture that the quality of orbit reconstruction is fundamentally limited by the accuracy of the force computation in the original simulation, which is inherently approximate in tree codes (see \citealt{Dehnen2014} for a discussion of force errors in conventional $N$-body codes).

\subsubsection{Dependence on the sampling interval}

\begin{figure}
\includegraphics{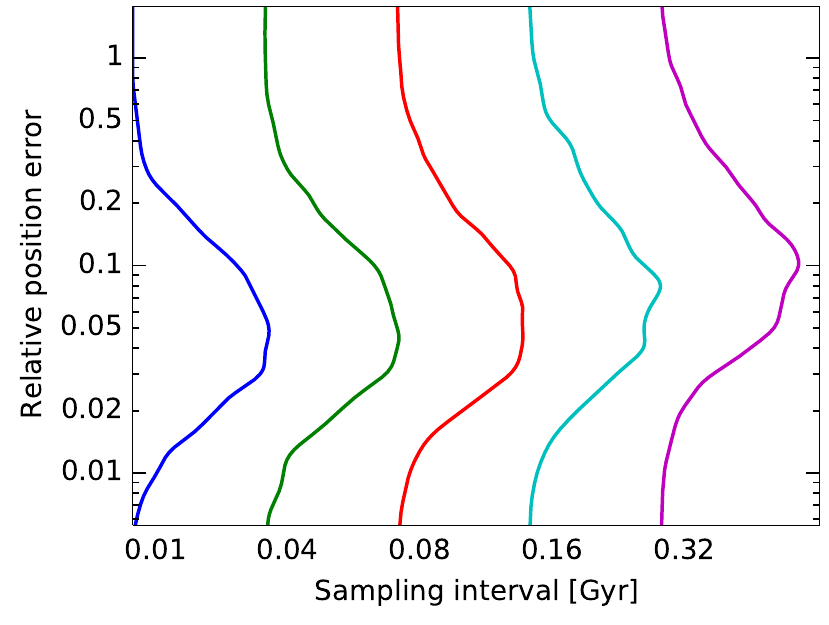}
\caption{Errors in reconstructed particle trajectories as functions of the sampling interval for the potential. The curves show the distribution of relative position errors after 2 Gyr, as in Figure~\ref{fig:violin} (here we consider only the Spline method with $l_\mathrm{max}=10$). The accuracy begins to deteriorate once this interval exceeds $\sim 0.1$~Gyr (although the bulk of orbits are not affected much, a tail of high relative errors becomes more prominent).
}  \label{fig:sampling}
\end{figure}

The $N$-body snapshots in our baseline scenario were stored rather frequently -- with a sampling interval of only 10~Myr. We now explore how the error $\mathcal E$ depends on this interval. Due to the additional smoothing of the halo centre-of-mass position, the spatially uniform acceleration associated with the non-inertial frame (Eq.~\ref{eq:force}) varies rather slowly in time. Fig.~\ref{fig:sampling} demonstrates that a tenfold increase of the snapshot spacing (to 100~Myr) does not significantly affect the error budget, but larger values start to deteriorate the accuracy, especially for a small fraction of orbits with already high relative errors. We find that using linear interpolation of force between two consecutive snapshots brings virtually no improvement compared to just taking the force from the nearest snapshot in time.

\begin{figure*}
    \centering
    \includegraphics[width=0.84\textwidth]{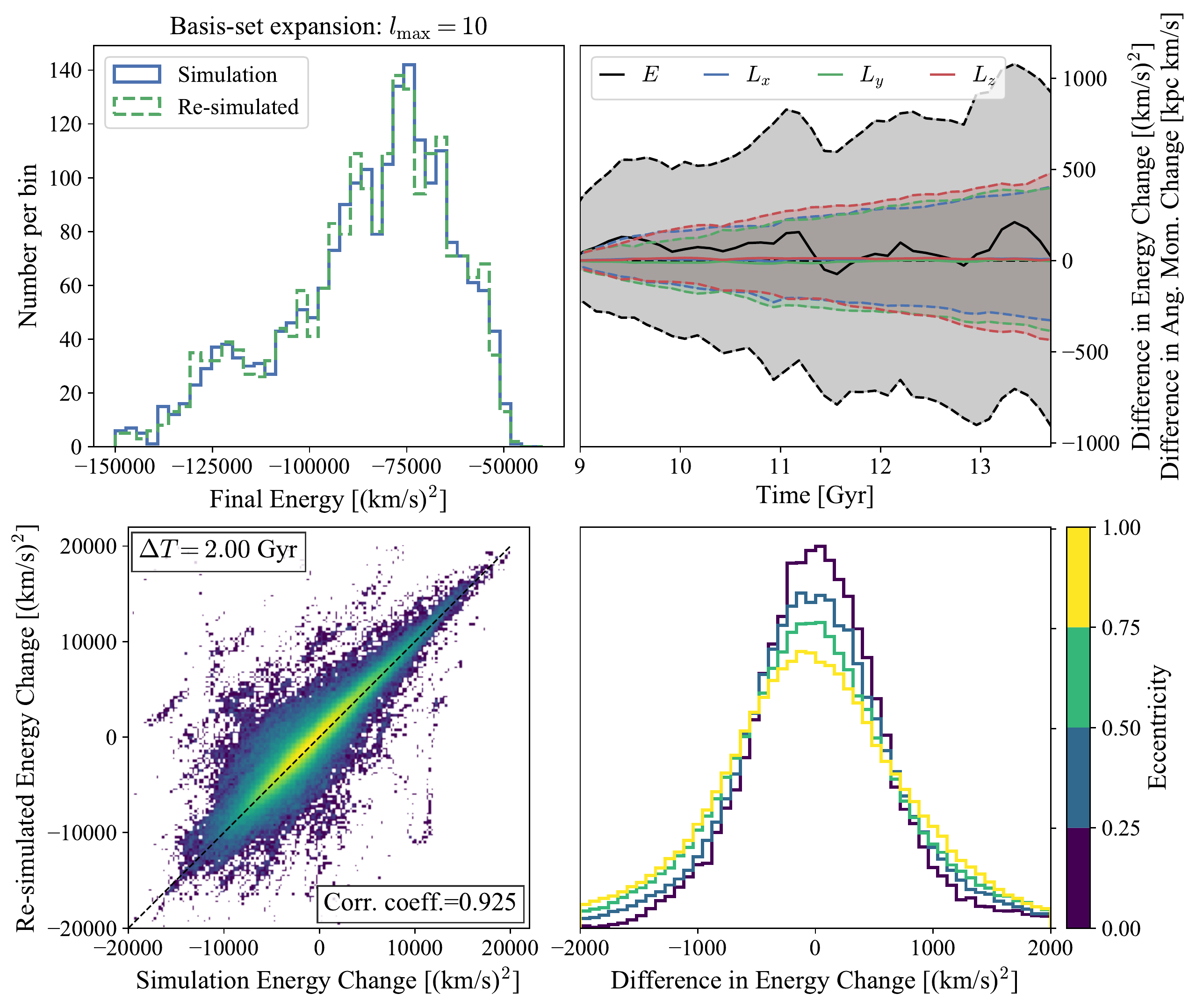}
    \caption{Evolution of energy and angular momentum for a sample of $\sim2000$ orbits: the top right panel shows the distributions of the final energies from the simulation (solid blue) and the basis-set expansion (dashed green). The top right panel shows the evolution of the difference in change in energy (black) and the components of the angular momentum (colours: blue $x$, green $y$, red $z$) between simulation and basis-set expansion. The median and $\pm1\sigma$ over orbits are shown. The bottom left panel shows the log-density of the energy changes in the simulation vs. basis-set expansion over $2\,\mathrm{Gyr}$ time intervals for all particles. The bottom right panel shows the difference in the energy change over $2\,\mathrm{Gyr}$ time intervals for orbits separated into quartiles of eccentricity. The more eccentric orbits are more poorly reproduced.}
    \label{fig:energy_change}
\end{figure*}

\subsubsection{Quality of Orbit Reproductions}

In addition to the error measure $\mathcal{E}$ useful for comparing the quality of different potential approximations, we can also inspect the overall success of our potential expansion methods through inspection of approximate integrals of motion. Despite the asphericity and time-dependence of the potential, the energy and angular momentum are still useful quantities for summarising a given orbit. In particular, we can check the quality of the orbit recovery by inspecting how well changes in these quantities are reproduced for our sample of orbits \citep{lowing2011}. In Fig.~\ref{fig:energy_change} we show some summary statistics for the changes in the integrals for our orbit sample using the $\lmax=10$ basis-function expansion. As the zeropoint of the potential is not well defined, we choose to match the median potential of the expansion to the median potential of the simulation evaluated at the location of all the inspected particles at each timestep. We observe that the distributions of the energies of the orbits at the end of the simulation are very satisfactorily recovered. The median of both the differences in the energy changes  and the difference in the angular momentum changes lie around zero at all times with a spread that grows steadily over time such that the dispersion is a few per cent in energy and a few tens of per cent in the components of angular momentum. An alternative way of displaying this information is to look at the energy changes over all $2\,\mathrm{Gyr}$ for all orbits (each orbit contributes multiple values). The majority of orbits lie along the one-to-one line with a small fraction forming clumps far off the line. The most common cause of this is subhalo scattering in the simulation. Finally, we split the difference in energy changes by orbital eccentricity (defined simply as $[\mathrm{max}(r)-\mathrm{min}(r)]/[\mathrm{max}(r)+\mathrm{min}(r)]$) and find that there is a weak trend for higher eccentricity orbits to be more poorly reproduced. These high eccentricity orbits are naturally more sensitive to successful reproduction of the potential over a wide range of radii, in particular the inner regions.

\subsection{Computational Cost}

Our previous discussion has focused on the accuracy of orbit reproduction for the basis expansion and spline expansion without any reference to the computational efficiency of the approaches. As we have demonstrated that both methods produce very similar results at similar order of expansion, it is then natural to ask which method is computationally cheaper. We concentrate on the evaluation costs as opposed to the setup costs: both methods require significant and comparable one-time upfront costs to find either sets of coefficients or spline fits. However, with these in place, a single evaluation of the potential is swift. 

\begin{figure}
    \centering
    \includegraphics{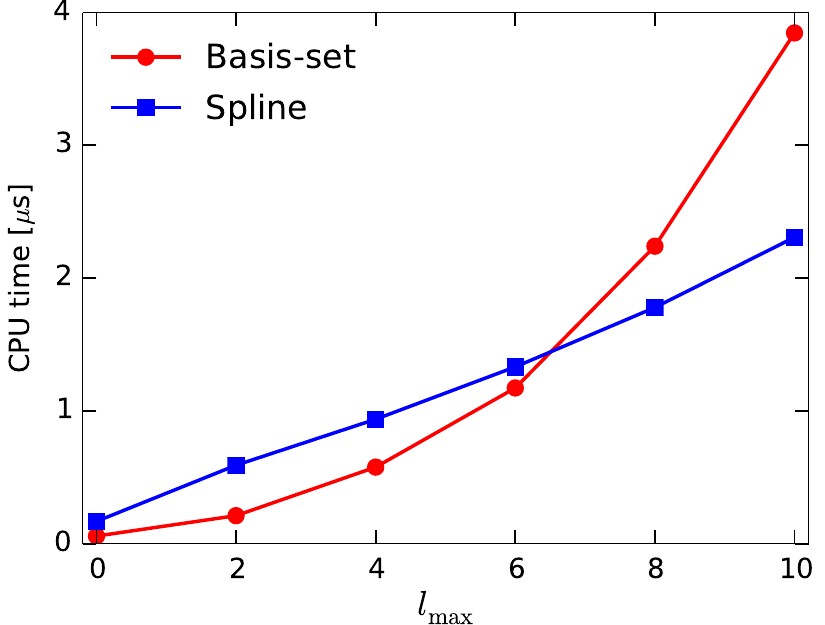}
    \caption{Cost of a single force evaluation as a function of the order of expansion: red circles -- basis-set expansion with $n_\mathrm{max} = 2\,l_\mathrm{max}+2$, blue boxes -- spline-interpolated multipole expansion. The latter scales nearly linearly with the number of Fourier terms, while the former scales cubically, but is cheaper for low orders. 
    In practice, when simulating a self-gravitating system embedded in an external potential, these costs are further amortized by the need to compute the inter-particle forces, which takes another $\sim 1-2\,\mu\mathrm s$ per particle in \textsc{gyrfalcON}.}
    \label{fig:cpu}
\end{figure}

Fig.~\ref{fig:cpu} shows the cost of a single force evaluation using each method with varying order of expansion. Force evaluation using the basis expansion scales approximately cubically with $\lmax$ as we require $(\nmax+1)(\lmax+1)(2\lmax+1)$ basis function evaluations and we have imposed $\nmax=2\lmax+2$. On the other hand, the spline expansion method (for $\lmax>2$) requires summing the 2d interpolated $(r,\theta)$ potential contribution $B_m(r,\theta)$ from each azimuthal order $m$ so scales approximately as $\mathcal O(2\lmax+1) + \mathcal O(\log\nmax+\log\lmax)$ (the log terms corresponding to the bisection algorithm used to locate the grid segment, and in practice are completely negligible for realistic orders of expansion). This means that for large numbers of terms the spline expansion method will always be more efficient per force evaluation, but at low expansion orders the basis set method is quicker (at given accuracy), as the approximate shape of the halo is already represented by the lowest few terms.

In the time-dependent potential pre-computed at discrete moments of time, one has several options for obtaining the force at any intermediate time: (1) use the potential from the nearest moment of time; (2) compute the force in the two nearest potentials and interpolate between them; (3) construct an potential at the intermediate moment of time by interpolating the coefficients of expansion. Options (1) and (3) involve only one force evaluation, while (2) requires two such evaluations. As explained above, interpolation of coefficients is trivial for the basis set, so (3) is the preferred variant: due to the linearity of all operations, it gives the same result as option (2), but costs almost the same as (1). However, option (3) is not practical for the spline representation, and we find that in our tests, the accuracy of options (1) and (2) are almost identical, so we prefer (1). Note, however, that if the potential is significantly varying between snapshots, time interpolation may become necessary, so the basis set method has an advantage. In more complex applications, we may have to evaluate the self-gravity of a re-simulated system (typically via a tree code) which using \textsc{gyrfalcON} scales as $\mathcal O(N)$ in the number of particles \citep{Dehnen2000} and takes $\sim1-2\mu\mathrm{s}$ per particle: a similar computational cost to the expansion methods.

\section{Applications}
\label{sec:app}

We now turn to some applications of the methods. The power of the presented approach is the ability to re-simulate sub-components or even insert new structures or features within the simulation without the significant cost of having to re-run the entire simulation. Provided any additions we make are of sufficiently low mass to essentially act as tracers in the simulations, the large-scale potential will be unchanged, and the potential expansion methods can be used to capture the details of the simulation down to any particular scale. As discussed previously, small scale structures such as dense subhaloes are not captured by the methods. If desired, they can be added as additional components in any re-simulation.

\cite{lowing2011} illustrated the power of this approach by simulating the disruption of a self-gravitating satellite galaxy within a Milky-Way-like halo. The large-scale potential field computed using a Hernquist-Ostriker expansion then acts as a tidal field on the satellite, and variation of initial conditions, such as orbit or internal structure, can be rapidly investigated.

We will investigate two simple applications of the approach that exploit the ability to rapidly re-simulate test particle orbits within a realistic time-dependent Milky-Way-like potential. The first is a brief investigation into the importance of a time-dependent potential on the interpretation of the orbits of the Milky Way's dwarf spheroidal galaxies and outer globular clusters. The second is an analysis of the dispersal of planes of satellites due to the time evolution of the host potential.

\subsection{Dwarf spheroidal and globular cluster orbits}

\begin{figure*}
  \centering
  \includegraphics[width=\textwidth]{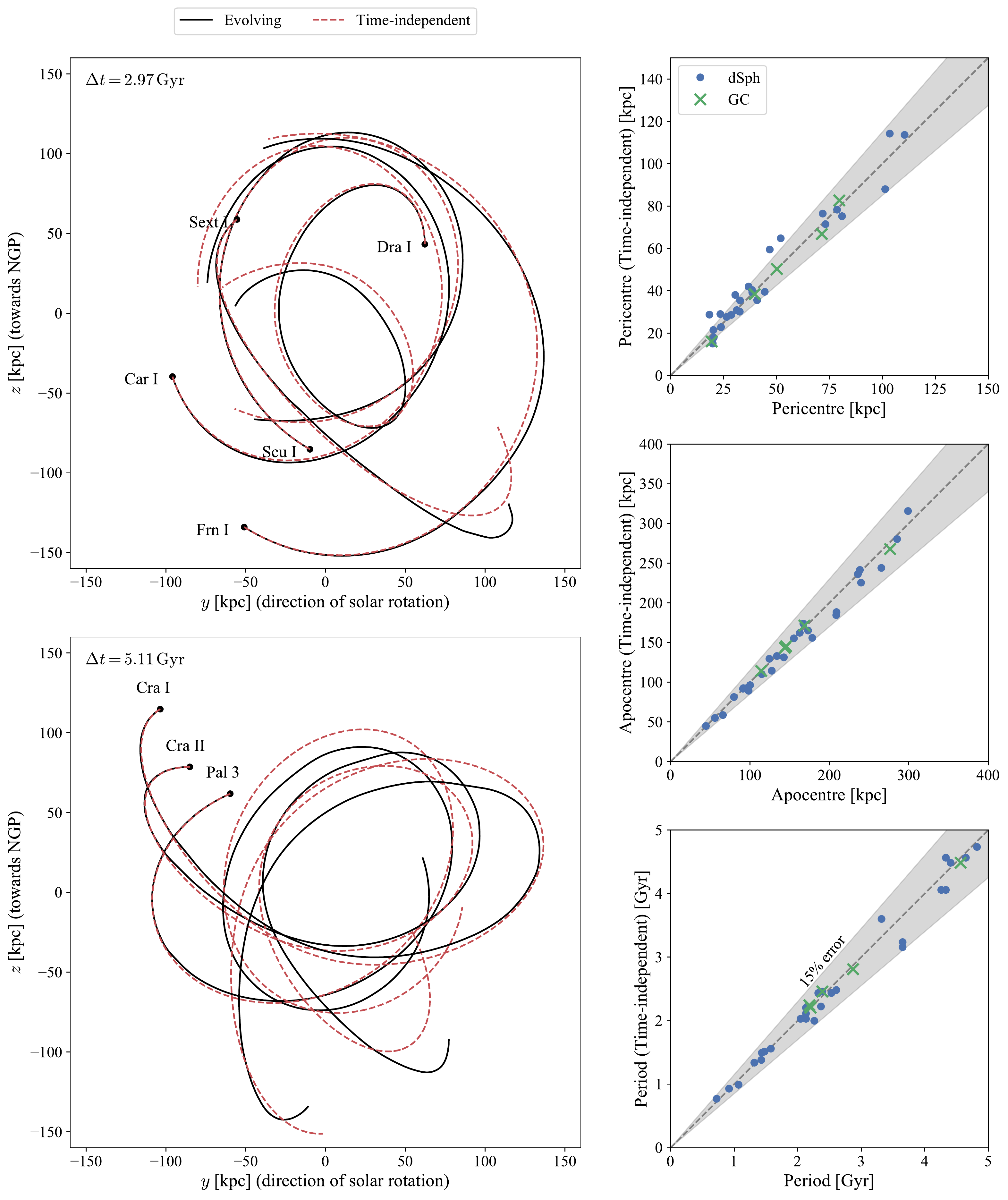}
  \caption{
  Dwarf spheroidal and outer globular cluster orbits within a time-evolving potential. The left panels shows two groups of orbits run backwards for $2.97\,\mathrm{Gyr}$ and $5.11\,\mathrm{Gyr}$ respectively in a fixed potential (red dashed) or full time-evolving potential (solid black). 
  The right three panels show the pericentres, apocentres and periods of a broader sample of dSphs (blue dots) and outer GCs (green crosses) in the full potential ($x$-axis) and time-independent ($y$-axis). 
  }
  \label{fig:dsph_gc}
\end{figure*}

The arrival of the Gaia data \citep{Gaia2016,Gaia2018} has allowed the 6D phase space coordinates of many dwarf spheroidal galaxies (dSphs) and globular clusters (GCs) to be measured for the first time or with significantly improved precision \citep{HelmiGaia}. Knowing the orbits of these objects allows assessment of the impact of tidal effects on their dynamical structure and in the case of dwarf spheroidal galaxies can have an impact on their interpretation within $\Lambda$CDM cosmology
\citep[e.g.][]{SandersCrater}. Uncertainty on the orbits of the dSphs and GCs is driven primarily by measurement errors and uncertainty on the present day potential of the Milky Way. However, both of these uncertainties can in theory be decreased with improved observations. A more fundamental uncertainty is the unknown time-dependence of the Milky Way's potential. Although it may be possible to infer likely time evolution of the Milky Way, such an observation can never be directly performed, meaning there is a limit to which we can know the past orbits of the dSphs and GCs. Here, we will assess the importance of this limitation by investigating the orbital properties of the known dSphs and outer halo GCs in a time-evolving versus time-independent potential.

For the GCs, we take the position, distance, proper motion and line-of-sight velocity data compiled by \cite{BaumgardtGCs}. For the dSphs, we use the positions, distances and line-of-sight velocities from the catalogue of \cite{McConnachie2012} (updated 25 Oct 2019). We complement this list with proper motions measured primarily from \cite{Fritz2018} and with additions from \cite{Massari2018} for Boo III and Sgr II, \cite{Torrealba2019} for Ant II, \cite{Pace2019} for Ret III, Phe II, Col I and Gru II, and \cite{Simon2020} for Tuc IV and Tuc V. We assume a solar location of $(R_0,z_0)=(8.18,0.02)\,\mathrm{kpc}$ and motion of $(11.1,247.3,7.25)\,\mathrm{km\,s}^{-1}$ \citep{Gravity,BennettBovy,ReidBrunthaler2004,Schoenrich}. For each object, we integrate the orbit backwards in the time-evolving potential of the halo described in Section~\ref{sec:halo_desc} and also in a fixed potential using only the final snapshot. We utilise the spline expansion method with $\lmax=8$ although the results would be unchanged if we used the basis expansion method. In the time-independent potential, we also set the acceleration of the halo centre to zero at all times.

We display the results of our procedure in Fig.~\ref{fig:dsph_gc}. Two sets of sample orbits are shown: five of the classical dwarf spheroidals Fornax, Sculptor, Carina, Sextans and Draco  which orbit in a similar plane, and the dwarf spheroidal Crater II and two globular clusters, Crater and Pal 3 (Crater's nature as a globular cluster as opposed to a dwarf spheroidal is uncertain). 
We find that although the time-independent orbits deviate from those integrated in the time-evolving potential, there is no obvious bias present. 
To quantify this, we have computed the pericentres, apocentres and periods within the time-dependent and time-independent potentials. We have removed orbits with pericentres $<15\,\mathrm{kpc}$ (as we don't have a disc potential our experiment is not reliable for these orbits) and apocentres $>400\,\mathrm{kpc}$, as well as those for which don't complete a peri/apocentric passage. We observe that the orbital properties lie along the one-to-one line with a $\sim15\,\mathrm{per\,cent}$ scatter in pericentre and $\sim5\,\mathrm{per\,cent}$ in apocentre and period with no obvious bias. It should be noted that we have only inspected a time interval of $\sim5\,\mathrm{Gyr}$ and over this time interval, the evolution of the inspected halo is quiet.

\jason{To connect to our investigations on orbit recovery from the previous section, we ran the orbits using different orders of expansion. If we take $l_\mathrm{max}=10$ as our most accurate orbit integration, the median $\mathcal{E}$ after 2$\,\mathrm{Gyr}$ for the inspected sample is $\sim(0.01,0.02,0.05)$ for $l_\mathrm{max}=(8,6,4)$ expansions respectively. This corresponds to $(1.3,2.5,2.9)\,\percent$ error in pericentre, $(0.4,1.0,1.9)\,\percent$ error in apocentre and $(0.7,1.6,1.8)\,\percent$ error in period. This verifies that the $l_\mathrm{max}=8$ expansion is more than adequate for our purposes here.
}

The amplitude of this scatter should be compared with both the scatter from the uncertainty in the potential at the current time and the uncertainty in the phase-space coordinates of the satellites. \cite{HelmiGaia} investigates the orbits of $10$ dSphs (mostly classical dwarfs) in three different potentials with similar circular velocity curves out to $\sim40\,\mathrm{kpc}$, but with a range of virial masses. The median scatter in the apocentres from the different potentials is $\sim5\,\percent$ whilst it is $\sim12\,\percent$ for the pericentres. The observational uncertainties from this sample produce uncertainties of $\sim8$ and $\sim20\,\percent$ in the apo- and pericentres respectively (in the median). \cite{Fritz2018} considers a broader sample of dwarfs with more ultrafaints which have poorer quality data. They consider two potentials with virial masses of $0.8$ and $1.6\times10^{12}M_\odot$. They find the different potentials produce variation of the apo- and pericentres of $\sim50\,\percent$ and $\sim10\,\percent$ respectively, whilst there is scatter of $\sim30\,\percent$ and $\sim40\,\percent$ from the observational uncertainties. Therefore, for the classical dSphs, the uncertainty arising from time-dependence of the potential is comparable to the uncertainty from the observational uncertainties and the uncertainty in the potential. However, for the ultrafaints, the time dependence of the potential is a significantly smaller uncertainty than that arising from the data quality. This will improve with future Gaia data releases.

In conclusion, we see that the time dependence of the Milky Way's potential is expected to produce $\sim15\,\mathrm{per\,cent}$ uncertainty in the gross orbital properties of the dwarf spheroidal galaxies and outer globular clusters. As our time-evolving halo has a modestly quiescent history, such uncertainties are likely a lower limit.

\subsection{Planes of satellites}

It was noted by \cite{LyndenBell1976} that many of the dSph galaxies known at that time lie in a plane which approximately follows the Magellanic stream. The ongoing discovery of dSphs has revealed many more potential plane members \citep{Pawlowski2015}. There is also supporting evidence from the intrinsic shapes and alignments of the satellites themselves~\citep{Sa17}. The plane has been hypothesised as arising from group or filamentary infall due to approximate alignment with large-scale structure \citep{Libeskind2015}, although both the observed thinness of the plane and the number of members has presented a challenge to $\Lambda$CDM theory due to the relatively low occurrence rate of such planes in cosmological simulations \citep[e.g.][]{Pawlowski2018,Shao2018}. This indicates that maybe the Milky Way's plane of satellites is due to a fortuitous alignment, although other planes of satellites have been suggested in both Andromeda and Centaurus A. The proper motions from Gaia have allowed better characterisation of the orbits of the dwarf spheroidals \citep{Fritz2018,Pawlowski2020}, which demonstrated that indeed some satellites (for instance, Sextans) lie momentarily within the plane, but do not orbit within it. However, the high quality proper motions from Gaia have confirmed the existence of the plane. Numerous studies have considered the growth of planes of satellites within fixed potentials \citep[e.g.,][]{Bowden2013,Erkal2016} or full cosmological simulations \citep{Shao2019}, generically finding that there is a tendency for planes to be longer lived if their orbital angular momentum lies near the long or short axes of the triaxial dark matter host halo. Furthermore, \cite{Shao2019} inspect hydrodynamical cosmological simulations and highlight the importance of the shape of the halo in reorienting and enhancing planes of satellites.

Here, we use our rapid re-simulation methods to assess the dispersal of planes of satellites due to the time dependence of the potential. We generate artificial planes of satellites by placing $50$ tracers at $50\,\mathrm{kpc}$ and random azimuthal angles. We choose a radius of $50\,\mathrm{kpc}$ such that a sufficient number of orbits is sampled in our time window. For each plane, we rotate by a random angle on the sphere and assign the tracers a tangential velocity in the plane equal to the local circular velocity. Finally, we scatter the velocities by a random dispersion $\sigma$ selected uniformly between $10$ and $40\,\mathrm{km\,s}^{-1}$ for each plane. Correspondingly, the tracers are scattered by $\sigma^2/|a_r|$ perpendicular to the plane for local radial acceleration $a_r$. The satellite orbits are then integrated in the full time-evolving potential with the accelerating centre, and in a potential fixed at that of the first snapshot with no acceleration of the centre. We use the $\lmax=8$ spline expansion method to compute the potential.

\begin{figure*}
  \centering
  \includegraphics[width=.77\textwidth]{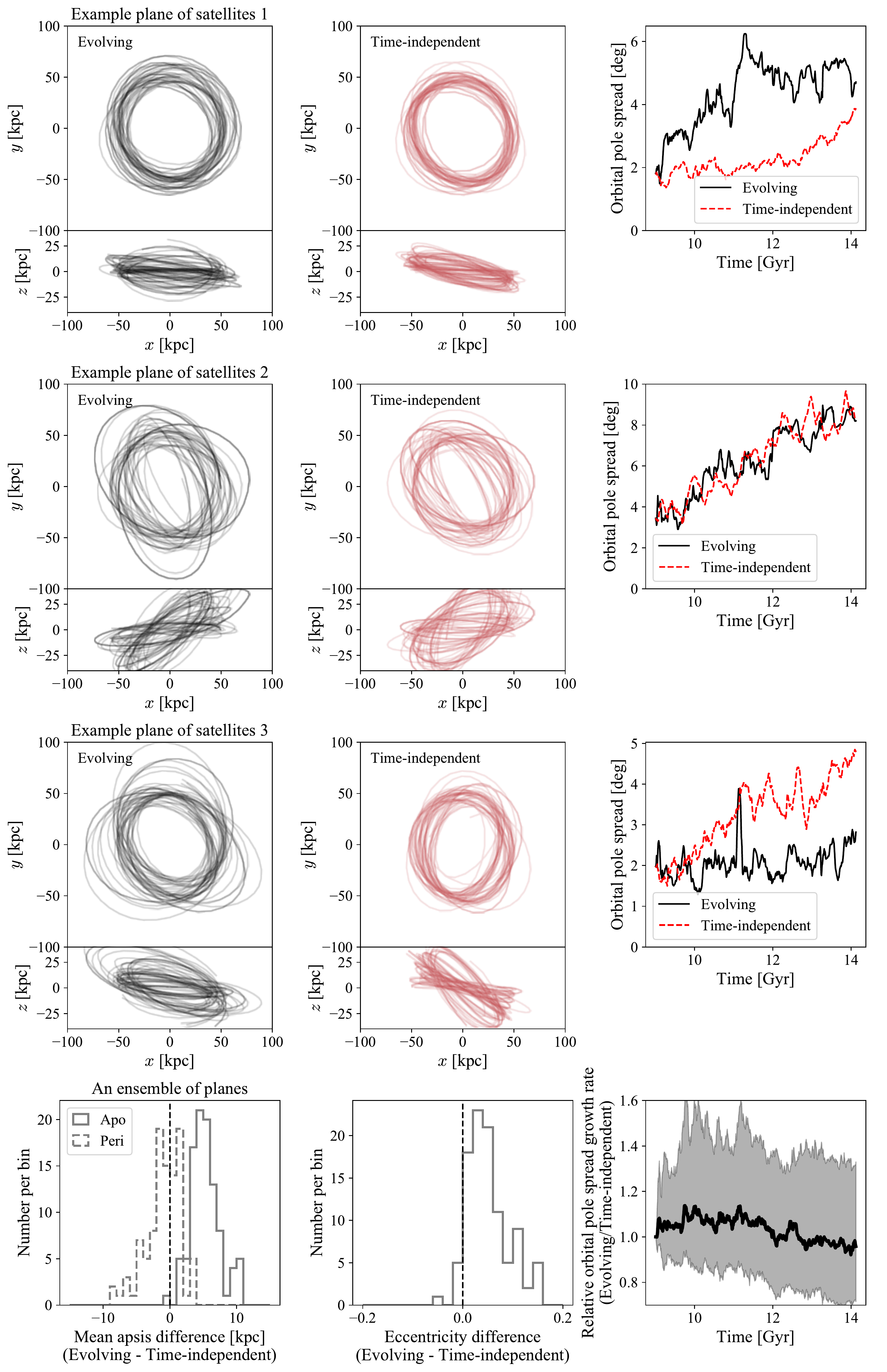}
  \caption{Planes of satellites within time-evolving potentials: in the top three rows, we show the properties of  example planes of satellites within a time-evolving potential (black) and the time-independent potential (red). The potentials are identical at the initial time. The $(x,y,z)$ coordinate system is chosen such that $z$ aligns with the angular momentum vector at the initial time. In the right panels, we show the spread in the orbital angular momentum pole direction. The three examples are discussed fully in the text. In the bottom row we show statistics for an ensemble of $100$ planes each with $50$ members. We show the difference in pericentre and apocentre in the bottom left, the difference in eccentricity in the middle, and the ensemble median with $\pm1\sigma$ bracket ratio of the final-to-initial orbital pole direction spread ratio in the evolving to time-independent potential.}
  \label{fig:pos}
\end{figure*}

In Fig.~\ref{fig:pos}, we show the evolution of three example planes from our sample. We depict a `face-on' view of the planes using the initial angular momentum vector, and a side-on projection. A generic feature is the increased eccentricity of the orbits in the evolving case. The interesting measure for the survival of the planes is their thickness over time. To quantify this, we compute half the difference between the $84$th and $16$th percentile of the angle between each tracer's angular momentum vector and the median angular momentum vector (the spread in orbital angular momentum). We see that the three examples chosen exhibit different behaviours: for one the evolving potential produces a larger dispersal of the plane than the time-independent, for another the opposite is true and for the final one the dispersal with time is approximately the same. A potential cause for these differences is due to the planes' alignments with respect to the halo axes. To investigate this, we compute the dot-product of the median orbital angular momentum of the satellite plane with the minor axis of the halo within $60\,\mathrm{kpc}$. For Example 3 (where the time-independent growth is greater than the time-dependent), we find the orbital angular momentum is well-aligned with the minor axis leading to slow initial growth. However, as the halo tumbles in the evolving potential the plane remains closely aligned to the minor axis and in fact is more aligned at the final time. This leads to the plane in the evolving potential broadening more slowly than in the time-independent. In Example 1 (where the evolving growth is larger than time-independent), the orbital angular momentum starts nearly aligned with the major axis but as the halo tumbles the orbital angular momentum becomes more misaligned with the major axis. As found by  \citet{Bowden2013} and \citet{Erkal2016}, alignment with the major axis leads to slower dispersal of the plane as seen in the time-independent case. Finally, Example 2 (where both time-independent and evolving growth are similar) is more complicated, as the orbital angular momentum starts at an intermediate angle with respect to both major and minor axis. In the evolving potential, it becomes more aligned with the major axis at the final time but in general it appears for these intermediate values the dispersal is very similar for both evolving and time-independent.

We inspect the statistics of our sample of planes in the bottom panels of Fig.~\ref{fig:pos}, which presents the difference in the pericentre, apocentre and eccentricty distributions. We find on average the mean apocentres and hence eccentricities of the plane members increase in the time-evolving potential, whilst the mean pericentres remain approximately fixed (as shown in the three example planes). Despite the change in the in-plane motions of the plane members, we find on average essentially no difference between the orbital pole spread in the evolving and time-independent potentials. It appears we have near equal numbers of the three example planes shown. This is quantified by computing the ratio of the final to initial orbital pole spread (the growth of the plane thickness, which we find is quite insensitive to the dispersion of the plane) and then plotting the ratio of this quantity in the evolving to time-independent potentials to measure the relative growth rates. We see from Fig.~\ref{fig:pos} that the growth rates in the time-independent and evolving cases are similar, with the weak signal of the time-independent potential producing slightly broader planes than the evolving potential. This echoes the conclusion of \cite{Shao2019} who found that the final properties of planes of satellites in a hydrodynamical cosmological simulation were more closely linked to the late-time properties of the host halo than the infall properties, and actually the evolution in the host halo could enhance the plane of satellites signal. 

\jason{As a final test, we ran a small sample of model planes of satellites using different orders of expansion. Compared to the $l_\mathrm{max}=10$ expansion, the error in the spread of the planes was $\sim(7,10,15)\,\percent$ for $l_\mathrm{max}=(8,6,4)$ respectively. This indicates that using $l_\mathrm{max}=8$ produces robust results for our particular application.}

In conclusion, we have seen that the time-dependence of the potential can alter the dispersal of planes of satellites particularly when the halo tumbles with respect to the plane and can present different halo shapes to the orbits of the satellites. This leads to more or less growth if the plane becomes less or more aligned with the major or minor axis of the halo. This is situation specific and in general we don't find an increase in dispersal in evolving potentials compared to time-independent. If anything, there are hints of the opposite effect, as an evolving potential can `shepherd' a plane and keep it thin.

\section{Conclusions}

In recent years, there has been growing awareness that galaxies do not have simple shapes and are not in equilibrium. This has been driven by high resolution simulations -- for example, the shapes of the dark haloes in the Auriga project show twisting and this often correlates with recent accretion or merger events~\citep{Pr19}. Observational evidence for disequilibrium is abundant for the Milky Way galaxy. A prominent example is the impending encounter of the Large Magellanic Cloud with the Milky Way, which affects the dynamics of stellar streams~\citep{Go15,Erkal2018} and which distorts the structure of the dark halo by an induced response~\citep{GC19,Be19,Pe20,Er20}. This has stimulated renewed attention on basis function methods, which have the flexibility to reproduce very general, time-varying gravitational fields, whether for dark haloes~\citep{Be20,Cu20} or other Galactic components like bars~\citep{PKW,Pe16}. 

A calibration of the performance of different basis function expansions against static galaxy models has already been performed by \citet{Vasiliev2013}. Here, we have provided a similar comparison, but for the harder problem of time-evolving models. Suppose we are given snapshots of an $N$-body simulation. For each snapshot, we represent the gravity field by basis function expansions. Interpolating between the expansions at each snapshot gives us a description of the evolving gravity field. How do the $N$-body orbits compare to the reconstructed orbits using the basis function expansions? To answer this question, we introduce a new error measure, based on the fidelity of the reconstructions. For each orbit, we compute the relative position error after a single period. Given a sample of orbits, the median and the spread of relative position errors allow us to quantify the performance of different expansions.

We examined two basis function methods in detail. The first uses biorthonormal expansions to represent the radial variation of the density and potential. The most familiar example is the \citet{hernquist1992} expansion. However, there are other possibilities in the literature~\citep{zhao1996}, whilst recent work~\citep{LSE,Lilley2018b} has provided an abundance of further such expansions. At zeroth order, the expansions have different density slopes at the centre and the outer parts, raising the possibility that the expansion can be tailored for any numerical halo. The second uses splines, an idea developed by \citet{Vasiliev2013}. In its most recent manifestation in the \wyn{public} software package \textsc{Agama} ~\citep{Va19}, quintic splines are employed with nodal points on a logarithmic radial grid. In both cases, spherical harmonics describe the angular dependence of the potential and forces. This also implies that these methods work best for spheroidal systems that are not too flattened, since otherwise a prohibitively large number of angular terms would be needed to accurately resolve thin disc-like structures. In those cases, alternative approaches such as CylSpline method (also provided in \textsc{Agama}) may be preferred \citep{Wang20}.

\medskip

Our main conclusions are as follows:

\begin{enumerate}

 \item Interpolation between $N$-body snapshots requires careful attention to the changing acceleration of the reference frame. The simulation particles are not just subject to the forces due to the halo itself, but also those due to the large-scale structure exterior to the halo. We find that numerical computation of the acceleration of the halo centre for each snapshot, followed by interpolation, performs satisfactorily in our orbit reconstructions. Strictly speaking, this approach neglects any tidal effects on the scale of an individual halo, which may be important for orbits with large apocentric distances.
  \medskip
  
 \item As regards the variety of biorthonormal expansions, the orbit reconstructions seem largely immune to any particular choice. All the biorthonormal expansions are complete, so can in principle reproduce any smooth density, but we might have expected fewer terms are needed if the zeroth order model is appropriately chosen to mimic the properties of the numerical halo.  In fact, despite the greater flexibility afforded at zeroth order by the expansions of \cite{Lilley2018b}, when using more than a few radial and angular terms we have found no reasons to use anything other than the \citet{hernquist1992} expansion  -- at least for the halo that we studied in detail here, with its cusp roughly constant in time, $\gamma \approx 1$.
 The behaviour of the error measure versus the free parameter of the biorthonormal expansions is essentially flat once even a modest number of terms is used, as shown in Fig.~\ref{fig:choosing_parameters}. This is an important conclusion as the Hernquist-Ostriker expansion is computationally the simplest and fastest expansion. Much work on orbit reconstruction has already been done using it ~\citep[e.g.,][]{lowing2011,Ngan2016,Bo18}. \wyn{Note that some caution is still needed as \citet{kalapotharakos2008} argue that the zeroth order model in a basis function expansion should have the same singular behaviour at the centre as the simulation. Our results do not directly address this point, as both simulation and zeroth order models have the same $1/r$ central density cusp.}
 
 \medskip  
      
 \item The coefficients in any biorthonormal expansions require tail corrections to avoid numerical artefacts caused by the edge, or the finite truncation radius, of numerical haloes. The corrections are important for low order expansions or for orbits that pass close to, or outside, the edge.
 \medskip 
 
 \item The spline and biorthonormal basis function methods are very comparable in terms of accuracy, and there is no compelling reason to prefer one over the other -- provided a reasonable number of terms are used. Our $N$-body halo has a minor to major axis ratio that varies from 0.5 to 0.8 over a Galactocentric range of 200 kpc in radius. 
 We find that the order of angular expansion $\lmax=6$ and the number of radial terms $\nmax \approx 15-20$ are sufficient for our problem. For both methods, the performance of individual orbit reconstructions in terms of pericentres, apocentres and eccentricities are normally fine over many orbital periods, but errors in the phase do gradually accumulate. This conclusion is evident from Figs.~\ref{fig:error} and ~\ref{fig:violin}. Longer period orbit in general tend to be less well reconstructed. In part, this is just a consequence of the fact that the relative position error is measured over a longer time for such orbits. However, both the effects of tidal forces and subhaloes are more important for larger apocentric orbits. A small number of orbits are poorly reconstructed, and this is usually due to scattering by subhaloes in the original simulation. The basis function reconstructions at the orders which we consider here of course only aim to reproduce the smooth underlying halo and do not account for small-scale substructure.

\medskip
\item  The computational costs of the spline and basis function methods are similar although force evaluation scales differently with the number of terms in the expansion (cubic for the basis function expansion and linear for the spline expansion). For the re-simulation of time-dependent systems, the basis function expansion is particularly efficient as the coefficients can be interpolated instead of the forces leading to fewer force evaluations.

\end{enumerate}

To demonstrate the power and use of the presented machinery, we investigated two simple applications. First, we inspected the impact of a time-dependent potential on the orbits of the Milky Way's population of satellites. We found that assuming a time-independent potential over the last $\sim5\,\mathrm{Gyr}$ produces an uncertainty in the orbital properties (pericentre, apocentre and period) of about $15\,\percent$. This is comparable to the uncertainty arising from the unknown Milky Way potential at the current time and that arising from observational errors for the classical dwarf galaxies. Secondly, we investigated the dispersal of planes of satellites injected into the halo, finding that the time-dependence of the halo potential does not make planes broader. The more significant effect is how the plane is aligned with respect to the triaxial halo, which in a time-dependent tumbling halo can change, leading to an increase \emph{or} decrease of the dispersal.

We believe that these simple applications have only scratched the surface of what is possible with basis expansion methods. For example, the Large Magellanic Cloud affects structures such as streams within the Milky Way Galaxy in at least three ways -- the time-dependent tides, the dark matter wake in the Milky Way halo in response to the mass of the Cloud and the motion of the Milky Way's centre of mass. The interplay between these three dynamical effects, all of which are known to be important, provides a superlative example for the use of basis function expansion methods. We therefore anticipate widespread applicability of these methods for the study of the substructure within, and evolution of, haloes -- both in the Milky Way galaxy and beyond.

\wyn{For those mainly interested in applications, it is useful to end with a quick \lq How-To'. We remain neutral as to the choice between splines and biorthonormal basis functions.} \nwyn{As starting points, the non-expert will find the spline expansion implemented in the publicly available~\textsc{Agama} software package~\citep{Va19}, whilst the Hernquist-Ostriker expansion is implemented in the publicly available \textsc{galpy} package~\citep{Bo15} or in \textsc{gyrfalcON} through the NEMO stellar dynamics toolkit~\citep[e.g.,][]{Mc07}.} Amongst the biorthornormal expansions, the Hernquist-Ostriker expansion is usually hard to beat, although other choices are available for different inner density slopes. Provided the halo is not strongly flattened, then a choice of $\approx 15$ radial and $\approx6$ angular terms is a good foothold to begin the problem. For comparison, \citet{lowing2011} and \citet{Ngan2015} used between 10 and 20 radial and angular terms. The spacing of the time snapshots depends on the violence of the accretion history of the halo, but $\approx$ 0.1 Gyr snapshots worked well here and in \citet{lowing2011}. To obtain high quality orbit reconstructions, it is very important to take into account the acceleration of the halo centre. This is best found by constructing a smoothed trajectory of the halo centre before numerically differentiating. It is an essential step in obtaining good results.

\section*{Acknowledgments}
JLS and EL acknowledge the support of the Royal Society and STFC respectively. EV is supported on the STFC Consolidated Grant awarded to the Institute of Astronomy. We also thank Matt Auger and members of the Institute of Astronomy Streams group for discussions and comments as this work was in progress. This research was supported in part by the National Science Foundation under Grant No. NSF PHY-1748958. This work has made use of data from the European Space Agency (ESA) mission
{\it Gaia} (\url{https://www.cosmos.esa.int/gaia}), processed by the {\it Gaia}
Data Processing and Analysis Consortium (DPAC,
\url{https://www.cosmos.esa.int/web/gaia/dpac/consortium}). Funding for the DPAC
has been provided by national institutions, in particular the institutions
participating in the {\it Gaia} Multilateral Agreement.

\section*{Data Availability}
The simulation data underlying this article will be shared on reasonable request to the corresponding author. All other data used are in the public domain and taken from 
\cite{McConnachie2012},
\cite{Fritz2018},
\cite{Massari2018},
\cite{Torrealba2019} and
\cite{Simon2020}.


\bibliographystyle{mnras}
\bibliography{bibliography.bib} 

\begin{thebibliography}{}
\makeatletter
\relax
\def\mn@urlcharsother{\let\do\@makeother \do\$\do\&\do\#\do\^\do\_\do\%\do\~}
\def\mn@doi{\begingroup\mn@urlcharsother \@ifnextchar [ {\mn@doi@}
  {\mn@doi@[]}}
\def\mn@doi@[#1]#2{\def\@tempa{#1}\ifx\@tempa\@empty \href
  {http://dx.doi.org/#2} {doi:#2}\else \href {http://dx.doi.org/#2} {#1}\fi
  \endgroup}
\def\mn@eprint#1#2{\mn@eprint@#1:#2::\@nil}
\def\mn@eprint@arXiv#1{\href {http://arxiv.org/abs/#1} {{\tt arXiv:#1}}}
\def\mn@eprint@dblp#1{\href {http://dblp.uni-trier.de/rec/bibtex/#1.xml}
  {dblp:#1}}
\def\mn@eprint@#1:#2:#3:#4\@nil{\def\@tempa {#1}\def\@tempb {#2}\def\@tempc
  {#3}\ifx \@tempc \@empty \let \@tempc \@tempb \let \@tempb \@tempa \fi \ifx
  \@tempb \@empty \def\@tempb {arXiv}\fi \@ifundefined
  {mn@eprint@\@tempb}{\@tempb:\@tempc}{\expandafter \expandafter \csname
  mn@eprint@\@tempb\endcsname \expandafter{\@tempc}}}

\bibitem[\protect\citeauthoryear{{Aarseth}}{{Aarseth}}{1967}]{Aarseth67}
{Aarseth} S.,  1967, in Les Nouvelles M\'ethodes de la Dynamique Stellaire.
  p.~47

\bibitem[\protect\citeauthoryear{{Baumgardt}, {Hilker}, {Sollima}  \&
  {Bellini}}{{Baumgardt} et~al.}{2019}]{BaumgardtGCs}
{Baumgardt} H.,  {Hilker} M.,  {Sollima} A.,   {Bellini} A.,  2019, \mn@doi
  [\mnras] {10.1093/mnras/sty2997}, \href
  {https://ui.adsabs.harvard.edu/abs/2019MNRAS.482.5138B} {482, 5138}

\bibitem[\protect\citeauthoryear{{Behroozi}, {Wechsler}  \& {Wu}}{{Behroozi}
  et~al.}{2013}]{rockstar}
{Behroozi} P.~S.,  {Wechsler} R.~H.,   {Wu} H.-Y.,  2013, \mn@doi [\apj]
  {10.1088/0004-637X/762/2/109}, \href
  {http://adsabs.harvard.edu/abs/2013ApJ...762..109B} {762, 109}

\bibitem[\protect\citeauthoryear{{Belokurov} et~al.,}{{Belokurov}
  et~al.}{2014}]{Be14}
{Belokurov} V.,  et~al., 2014, \mn@doi [\mnras] {10.1093/mnras/stt1862}, \href
  {https://ui.adsabs.harvard.edu/abs/2014MNRAS.437..116B} {437, 116}

\bibitem[\protect\citeauthoryear{{Belokurov}, {Deason}, {Erkal}, {Koposov},
  {Carballo-Bello}, {Smith}, {Jethwa}  \& {Navarrete}}{{Belokurov}
  et~al.}{2019}]{Be19}
{Belokurov} V.,  {Deason} A.~J.,  {Erkal} D.,  {Koposov} S.~E.,
  {Carballo-Bello} J.~A.,  {Smith} M.~C.,  {Jethwa} P.,   {Navarrete} C.,
  2019, \mn@doi [\mnras] {10.1093/mnrasl/slz101}, \href
  {https://ui.adsabs.harvard.edu/abs/2019MNRAS.488L..47B} {488, L47}

\bibitem[\protect\citeauthoryear{{Bennett} \& {Bovy}}{{Bennett} \&
  {Bovy}}{2019}]{BennettBovy}
{Bennett} M.,  {Bovy} J.,  2019, \mn@doi [\mnras] {10.1093/mnras/sty2813},
  \href {https://ui.adsabs.harvard.edu/abs/2019MNRAS.482.1417B} {482, 1417}

\bibitem[\protect\citeauthoryear{{Besla} \& {Garavito-Camargo}}{{Besla} \&
  {Garavito-Camargo}}{2020}]{Be20}
{Besla} G.,  {Garavito-Camargo} N.,  2020, in {Valluri} M.,  {Sellwood} J.~A.,
  eds,  IAU Symposium Vol. 353, IAU Symposium. pp 123--127,
  \mn@doi{10.1017/S1743921319008494}

\bibitem[\protect\citeauthoryear{{Bonaca} \& {Hogg}}{{Bonaca} \&
  {Hogg}}{2018}]{Bo18}
{Bonaca} A.,  {Hogg} D.~W.,  2018, \mn@doi [\apj] {10.3847/1538-4357/aae4da},
  \href {https://ui.adsabs.harvard.edu/abs/2018ApJ...867..101B} {867, 101}

\bibitem[\protect\citeauthoryear{{Bovy}}{{Bovy}}{2015}]{Bo15}
{Bovy} J.,  2015, \mn@doi [\apjs] {10.1088/0067-0049/216/2/29}, \href
  {https://ui.adsabs.harvard.edu/abs/2015ApJS..216...29B} {216, 29}

\bibitem[\protect\citeauthoryear{{Bowden}, {Evans}  \& {Belokurov}}{{Bowden}
  et~al.}{2013}]{Bowden2013}
{Bowden} A.,  {Evans} N.~W.,   {Belokurov} V.,  2013, \mn@doi [\mnras]
  {10.1093/mnras/stt1253}, \href
  {https://ui.adsabs.harvard.edu/abs/2013MNRAS.435..928B} {435, 928}

\bibitem[\protect\citeauthoryear{{Boylan-Kolchin}, {Ma}  \&
  {Quataert}}{{Boylan-Kolchin} et~al.}{2008}]{Bo08}
{Boylan-Kolchin} M.,  {Ma} C.-P.,   {Quataert} E.,  2008, \mn@doi [\mnras]
  {10.1111/j.1365-2966.2007.12530.x}, \href
  {https://ui.adsabs.harvard.edu/abs/2008MNRAS.383...93B} {383, 93}

\bibitem[\protect\citeauthoryear{{Clutton-Brock}}{{Clutton-Brock}}{1972}]{cluttonbrock1972}
{Clutton-Brock} M.,  1972, \mn@doi [\apss] {10.1007/BF00643095}, \href
  {http://adsabs.harvard.edu/abs/1972Ap%26SS..16..101C} {16, 101}

\bibitem[\protect\citeauthoryear{{Clutton-Brock}}{{Clutton-Brock}}{1973}]{cluttonbrock1973}
{Clutton-Brock} M.,  1973, \mn@doi [\apss] {10.1007/BF00647652}, \href
  {http://adsabs.harvard.edu/abs/1973Ap%26SS..23...55C} {23, 55}

\bibitem[\protect\citeauthoryear{{Cunningham} et~al.,}{{Cunningham}
  et~al.}{2020}]{Cu20}
{Cunningham} E.~C.,  et~al., 2020, arXiv, \href
  {https://ui.adsabs.harvard.edu/abs/2020arXiv200608621C} {p. 2006.08621}

\bibitem[\protect\citeauthoryear{{Dehnen}}{{Dehnen}}{2000}]{Dehnen2000}
{Dehnen} W.,  2000, \mn@doi [\apjl] {10.1086/312724}, \href
  {https://ui.adsabs.harvard.edu/abs/2000ApJ...536L..39D} {536, L39}

\bibitem[\protect\citeauthoryear{{Dehnen}}{{Dehnen}}{2014}]{Dehnen2014}
{Dehnen} W.,  2014, \mn@doi [Computational Astrophysics and Cosmology]
  {10.1186/s40668-014-0001-7}, \href
  {http://adsabs.harvard.edu/abs/2014ComAC...1....1D} {1, 1}

\bibitem[\protect\citeauthoryear{{Dekel}, {Ishai}, {Dutton}  \&
  {Maccio}}{{Dekel} et~al.}{2017}]{Dekel2016}
{Dekel} A.,  {Ishai} G.,  {Dutton} A.~A.,   {Maccio} A.~V.,  2017, \mn@doi
  [\mnras] {10.1093/mnras/stx486}, \href
  {https://ui.adsabs.harvard.edu/abs/2017MNRAS.468.1005D} {468, 1005}

\bibitem[\protect\citeauthoryear{{Diemer} \& {Kravtsov}}{{Diemer} \&
  {Kravtsov}}{2014}]{Di2014}
{Diemer} B.,  {Kravtsov} A.~V.,  2014, \mn@doi [\apj]
  {10.1088/0004-637X/789/1/1}, \href
  {http://adsabs.harvard.edu/abs/2014ApJ...789....1D} {789, 1}

\bibitem[\protect\citeauthoryear{{Dubinski} \& {Carlberg}}{{Dubinski} \&
  {Carlberg}}{1991}]{Du91}
{Dubinski} J.,  {Carlberg} R.~G.,  1991, \mn@doi [\apj] {10.1086/170451}, \href
  {https://ui.adsabs.harvard.edu/abs/1991ApJ...378..496D} {378, 496}

\bibitem[\protect\citeauthoryear{{Einasto} \& {Haud}}{{Einasto} \&
  {Haud}}{1989}]{Ei89}
{Einasto} J.,  {Haud} U.,  1989, \aap, \href
  {https://ui.adsabs.harvard.edu/abs/1989A&A...223...89E} {223, 89}

\bibitem[\protect\citeauthoryear{Erkal, Sanders  \& Belokurov}{Erkal
  et~al.}{2016}]{Erkal2016}
Erkal D.,  Sanders J.~L.,   Belokurov V.,  2016, \mn@doi [\mnras]
  {10.1093/mnras/stw1400}, 461, 1590

\bibitem[\protect\citeauthoryear{{Erkal} et~al.,}{{Erkal}
  et~al.}{2019}]{Erkal2018}
{Erkal} D.,  et~al., 2019, \mn@doi [\mnras] {10.1093/mnras/stz1371}, \href
  {https://ui.adsabs.harvard.edu/abs/2019MNRAS.487.2685E} {487, 2685}

\bibitem[\protect\citeauthoryear{{Erkal}, {Belokurov}  \& {Parkin}}{{Erkal}
  et~al.}{2020}]{Er20}
{Erkal} D.,  {Belokurov} V.,   {Parkin} D.~L.,  2020, arXiv e-prints, \href
  {https://ui.adsabs.harvard.edu/abs/2020arXiv200111030E} {p. arXiv:2001.11030}

\bibitem[\protect\citeauthoryear{{Evans} \& {An}}{{Evans} \& {An}}{2006}]{Ev06}
{Evans} N.~W.,  {An} J.~H.,  2006, \mn@doi [\prd] {10.1103/PhysRevD.73.023524},
  \href {http://adsabs.harvard.edu/abs/2006PhRvD..73b3524E} {73, 023524}

\bibitem[\protect\citeauthoryear{{Fritz}, {Battaglia}, {Pawlowski},
  {Kallivayalil}, {van der Marel}, {Sohn}, {Brook}  \& {Besla}}{{Fritz}
  et~al.}{2018}]{Fritz2018}
{Fritz} T.~K.,  {Battaglia} G.,  {Pawlowski} M.~S.,  {Kallivayalil} N.,  {van
  der Marel} R.,  {Sohn} S.~T.,  {Brook} C.,   {Besla} G.,  2018, \mn@doi
  [\aap] {10.1051/0004-6361/201833343}, \href
  {https://ui.adsabs.harvard.edu/abs/2018A&A...619A.103F} {619, A103}

\bibitem[\protect\citeauthoryear{{Gaia Collaboration}}{{Gaia
  Collaboration}}{2016}]{Gaia2016}
{Gaia Collaboration} 2016, \mn@doi [\aap] {10.1051/0004-6361/201629272}, \href
  {https://ui.adsabs.harvard.edu/abs/2016A&A...595A...1G} {595, A1}

\bibitem[\protect\citeauthoryear{{Gaia Collaboration}}{{Gaia
  Collaboration}}{2018a}]{Gaia2018}
{Gaia Collaboration} 2018a, \mn@doi [\aap] {10.1051/0004-6361/201833051}, \href
  {https://ui.adsabs.harvard.edu/abs/2018A&A...616A...1G} {616, A1}

\bibitem[\protect\citeauthoryear{{Gaia Collaboration}}{{Gaia
  Collaboration}}{2018b}]{HelmiGaia}
{Gaia Collaboration} 2018b, \mn@doi [\aap] {10.1051/0004-6361/201832698}, \href
  {https://ui.adsabs.harvard.edu/abs/2018A&A...616A..12G} {616, A12}

\bibitem[\protect\citeauthoryear{{Garavito-Camargo}, {Besla}, {Laporte},
  {Johnston}, {G{\'o}mez}  \& {Watkins}}{{Garavito-Camargo}
  et~al.}{2019}]{GC19}
{Garavito-Camargo} N.,  {Besla} G.,  {Laporte} C. F.~P.,  {Johnston} K.~V.,
  {G{\'o}mez} F.~A.,   {Watkins} L.~L.,  2019, \mn@doi [\apj]
  {10.3847/1538-4357/ab32eb}, \href
  {https://ui.adsabs.harvard.edu/abs/2019ApJ...884...51G} {884, 51}

\bibitem[\protect\citeauthoryear{{G{\'o}mez}, {Besla}, {Carpintero},
  {Villalobos}, {O'Shea}  \& {Bell}}{{G{\'o}mez} et~al.}{2015}]{Go15}
{G{\'o}mez} F.~A.,  {Besla} G.,  {Carpintero} D.~D.,  {Villalobos} {\'A}.,
  {O'Shea} B.~W.,   {Bell} E.~F.,  2015, \mn@doi [\apj]
  {10.1088/0004-637X/802/2/128}, \href
  {https://ui.adsabs.harvard.edu/abs/2015ApJ...802..128G} {802, 128}

\bibitem[\protect\citeauthoryear{{Gravity Collaboration (Abuter et
  al.)}}{{Gravity Collaboration (Abuter et al.)}}{2019}]{Gravity}
{Gravity Collaboration (Abuter et al.)} 2019, \mn@doi [\aap]
  {10.1051/0004-6361/201935656}, \href
  {https://ui.adsabs.harvard.edu/abs/2019A&A...625L..10G} {625, L10}

\bibitem[\protect\citeauthoryear{{Hahn} \& {Abel}}{{Hahn} \&
  {Abel}}{2011}]{Ha11}
{Hahn} O.,  {Abel} T.,  2011, \mn@doi [\mnras]
  {10.1111/j.1365-2966.2011.18820.x}, \href
  {https://ui.adsabs.harvard.edu/abs/2011MNRAS.415.2101H} {415, 2101}

\bibitem[\protect\citeauthoryear{Hall}{Hall}{1983}]{Hall1983}
Hall P.,  1983, \mn@doi [Ann. Statist.] {10.1214/aos/1176346329}, 11, 1156

\bibitem[\protect\citeauthoryear{{Hernquist}}{{Hernquist}}{1990}]{Hernquist1990}
{Hernquist} L.,  1990, \mn@doi [\apj] {10.1086/168845}, \href
  {http://adsabs.harvard.edu/abs/1990ApJ...356..359H} {356, 359}

\bibitem[\protect\citeauthoryear{{Hernquist} \& {Ostriker}}{{Hernquist} \&
  {Ostriker}}{1992}]{hernquist1992}
{Hernquist} L.,  {Ostriker} J.~P.,  1992, \mn@doi [\apj] {10.1086/171025},
  \href {http://adsabs.harvard.edu/abs/1992ApJ...386..375H} {386, 375}

\bibitem[\protect\citeauthoryear{{Jing} \& {Suto}}{{Jing} \&
  {Suto}}{2002}]{Jing2002}
{Jing} Y.~P.,  {Suto} Y.,  2002, \mn@doi [\apj] {10.1086/341065}, \href
  {http://adsabs.harvard.edu/abs/2002ApJ...574..538J} {574, 538}

\bibitem[\protect\citeauthoryear{{Kalapotharakos}, {Efthymiopoulos}  \&
  {Voglis}}{{Kalapotharakos} et~al.}{2008}]{kalapotharakos2008}
{Kalapotharakos} C.,  {Efthymiopoulos} C.,   {Voglis} N.,  2008, \mn@doi
  [\mnras] {10.1111/j.1365-2966.2007.12417.x}, \href
  {http://adsabs.harvard.edu/abs/2008MNRAS.383..971K} {383, 971}

\bibitem[\protect\citeauthoryear{{Klypin}, {Kravtsov}, {Bullock}  \&
  {Primack}}{{Klypin} et~al.}{2001}]{Kl01}
{Klypin} A.,  {Kravtsov} A.~V.,  {Bullock} J.~S.,   {Primack} J.~R.,  2001,
  \mn@doi [\apj] {10.1086/321400}, \href
  {https://ui.adsabs.harvard.edu/abs/2001ApJ...554..903K} {554, 903}

\bibitem[\protect\citeauthoryear{{Laine} et~al.,}{{Laine} et~al.}{2018}]{Laine}
{Laine} S.,  et~al., 2018, arXiv, \href
  {https://ui.adsabs.harvard.edu/abs/2018arXiv181204897L} {p. 1812.04897}

\bibitem[\protect\citeauthoryear{{Law} \& {Majewski}}{{Law} \&
  {Majewski}}{2010}]{La10}
{Law} D.~R.,  {Majewski} S.~R.,  2010, \mn@doi [\apj]
  {10.1088/0004-637X/714/1/229}, \href
  {https://ui.adsabs.harvard.edu/abs/2010ApJ...714..229L} {714, 229}

\bibitem[\protect\citeauthoryear{{Libeskind}, {Hoffman}, {Tully}, {Courtois},
  {Pomar{\`e}de}, {Gottl{\"o}ber}  \& {Steinmetz}}{{Libeskind}
  et~al.}{2015}]{Libeskind2015}
{Libeskind} N.~I.,  {Hoffman} Y.,  {Tully} R.~B.,  {Courtois} H.~M.,
  {Pomar{\`e}de} D.,  {Gottl{\"o}ber} S.,   {Steinmetz} M.,  2015, \mn@doi
  [\mnras] {10.1093/mnras/stv1302}, \href
  {https://ui.adsabs.harvard.edu/abs/2015MNRAS.452.1052L} {452, 1052}

\bibitem[\protect\citeauthoryear{{Lilley}}{{Lilley}}{2020}]{Li20}
{Lilley} E.~J.,  2020, {PhD thesis}.
{Cambridge University}

\bibitem[\protect\citeauthoryear{{Lilley}, {Sanders}, {Evans}  \&
  {Erkal}}{{Lilley} et~al.}{2018a}]{LSE}
{Lilley} E.~J.,  {Sanders} J.~L.,  {Evans} N.~W.,   {Erkal} D.,  2018a, \mn@doi
  [\mnras] {10.1093/mnras/sty296}, \href
  {http://adsabs.harvard.edu/abs/2018MNRAS.476.2092L} {476, 2092}

\bibitem[\protect\citeauthoryear{{Lilley}, {Sanders}  \& {Evans}}{{Lilley}
  et~al.}{2018b}]{Lilley2018b}
{Lilley} E.~J.,  {Sanders} J.~L.,   {Evans} N.~W.,  2018b, \mn@doi [\mnras]
  {10.1093/mnras/sty1038}, \href
  {http://adsabs.harvard.edu/abs/2018MNRAS.478.1281L} {478, 1281}

\bibitem[\protect\citeauthoryear{{Lowing}, {Jenkins}, {Eke}  \&
  {Frenk}}{{Lowing} et~al.}{2011}]{lowing2011}
{Lowing} B.,  {Jenkins} A.,  {Eke} V.,   {Frenk} C.,  2011, \mn@doi [\mnras]
  {10.1111/j.1365-2966.2011.19222.x}, \href
  {http://adsabs.harvard.edu/abs/2011MNRAS.416.2697L} {416, 2697}

\bibitem[\protect\citeauthoryear{{Lynden-Bell}}{{Lynden-Bell}}{1976}]{LyndenBell1976}
{Lynden-Bell} D.,  1976, \mn@doi [\mnras] {10.1093/mnras/174.3.695}, \href
  {https://ui.adsabs.harvard.edu/abs/1976MNRAS.174..695L} {174, 695}

\bibitem[\protect\citeauthoryear{{Massari} \& {Helmi}}{{Massari} \&
  {Helmi}}{2018}]{Massari2018}
{Massari} D.,  {Helmi} A.,  2018, \mn@doi [\aap] {10.1051/0004-6361/201833367},
  \href {https://ui.adsabs.harvard.edu/abs/2018A&A...620A.155M} {620, A155}

\bibitem[\protect\citeauthoryear{{McConnachie}}{{McConnachie}}{2012}]{McConnachie2012}
{McConnachie} A.~W.,  2012, \mn@doi [\aj] {10.1088/0004-6256/144/1/4}, \href
  {https://ui.adsabs.harvard.edu/abs/2012AJ....144....4M} {144, 4}

\bibitem[\protect\citeauthoryear{{McGlynn}}{{McGlynn}}{1984}]{McGlynn84}
{McGlynn} T.~A.,  1984, \mn@doi [\apj] {10.1086/162072}, \href
  {https://ui.adsabs.harvard.edu/abs/1984ApJ...281...13M} {281, 13}

\bibitem[\protect\citeauthoryear{{McMillan} \& {Dehnen}}{{McMillan} \&
  {Dehnen}}{2007}]{Mc07}
{McMillan} P.~J.,  {Dehnen} W.,  2007, \mn@doi [\mnras]
  {10.1111/j.1365-2966.2007.11753.x}, \href
  {https://ui.adsabs.harvard.edu/abs/2007MNRAS.378..541M} {378, 541}

\bibitem[\protect\citeauthoryear{{Meiron}, {Li}, {Holley-Bockelmann}  \&
  {Spurzem}}{{Meiron} et~al.}{2014}]{Meiron14}
{Meiron} Y.,  {Li} B.,  {Holley-Bockelmann} K.,   {Spurzem} R.,  2014, \mn@doi
  [\apj] {10.1088/0004-637X/792/2/98}, \href
  {https://ui.adsabs.harvard.edu/abs/2014ApJ...792...98M} {792, 98}

\bibitem[\protect\citeauthoryear{{Merritt}, {Graham}, {Moore}, {Diemand}  \&
  {Terzi{\'c}}}{{Merritt} et~al.}{2006}]{Merritt06}
{Merritt} D.,  {Graham} A.~W.,  {Moore} B.,  {Diemand} J.,   {Terzi{\'c}} B.,
  2006, \mn@doi [\aj] {10.1086/508988}, \href
  {http://adsabs.harvard.edu/abs/2006AJ....132.2685M} {132, 2685}

\bibitem[\protect\citeauthoryear{{Mo}, {van den Bosch}  \& {White}}{{Mo}
  et~al.}{2010}]{MBW}
{Mo} H.,  {van den Bosch} F.~C.,   {White} S.,  2010, {Galaxy Formation and
  Evolution}.
{Cambridge University Press, Cambridge}

\bibitem[\protect\citeauthoryear{{Moore}, {Governato}, {Quinn}, {Stadel}  \&
  {Lake}}{{Moore} et~al.}{1998}]{Mo98}
{Moore} B.,  {Governato} F.,  {Quinn} T.,  {Stadel} J.,   {Lake} G.,  1998,
  \mn@doi [\apjl] {10.1086/311333}, \href
  {https://ui.adsabs.harvard.edu/abs/1998ApJ...499L...5M} {499, L5}

\bibitem[\protect\citeauthoryear{{Moore}, {Ghigna}, {Governato}, {Lake},
  {Quinn}, {Stadel}  \& {Tozzi}}{{Moore} et~al.}{1999}]{Mo99}
{Moore} B.,  {Ghigna} S.,  {Governato} F.,  {Lake} G.,  {Quinn} T.,  {Stadel}
  J.,   {Tozzi} P.,  1999, \mn@doi [\apjl] {10.1086/312287}, \href
  {https://ui.adsabs.harvard.edu/abs/1999ApJ...524L..19M} {524, L19}

\bibitem[\protect\citeauthoryear{{Navarro}, {Frenk}  \& {White}}{{Navarro}
  et~al.}{1997}]{NFW1997}
{Navarro} J.~F.,  {Frenk} C.~S.,   {White} S.~D.~M.,  1997, \mn@doi [\apj]
  {10.1086/304888}, \href {http://adsabs.harvard.edu/abs/1997ApJ...490..493N}
  {490, 493}

\bibitem[\protect\citeauthoryear{{Ngan}, {Bozek}, {Carlberg}, {Wyse}, {Szalay}
  \& {Madau}}{{Ngan} et~al.}{2015}]{Ngan2015}
{Ngan} W.,  {Bozek} B.,  {Carlberg} R.~G.,  {Wyse} R.~F.~G.,  {Szalay} A.~S.,
  {Madau} P.,  2015, \mn@doi [\apj] {10.1088/0004-637X/803/2/75}, \href
  {http://adsabs.harvard.edu/abs/2015ApJ...803...75N} {803, 75}

\bibitem[\protect\citeauthoryear{{Ngan}, {Carlberg}, {Bozek}, {Wyse}, {Szalay}
  \& {Madau}}{{Ngan} et~al.}{2016}]{Ngan2016}
{Ngan} W.,  {Carlberg} R.~G.,  {Bozek} B.,  {Wyse} R.~F.~G.,  {Szalay} A.~S.,
  {Madau} P.,  2016, \mn@doi [\apj] {10.3847/0004-637X/818/2/194}, \href
  {http://adsabs.harvard.edu/abs/2016ApJ...818..194N} {818, 194}

\bibitem[\protect\citeauthoryear{{O{\~n}orbe}, {Garrison-Kimmel}, {Maller},
  {Bullock}, {Rocha}  \& {Hahn}}{{O{\~n}orbe} et~al.}{2014}]{onorbe}
{O{\~n}orbe} J.,  {Garrison-Kimmel} S.,  {Maller} A.~H.,  {Bullock} J.~S.,
  {Rocha} M.,   {Hahn} O.,  2014, \mn@doi [\mnras] {10.1093/mnras/stt2020},
  \href {http://adsabs.harvard.edu/abs/2014MNRAS.437.1894O} {437, 1894}

\bibitem[\protect\citeauthoryear{{Pace} \& {Li}}{{Pace} \&
  {Li}}{2019}]{Pace2019}
{Pace} A.~B.,  {Li} T.~S.,  2019, \mn@doi [\apj] {10.3847/1538-4357/ab0aee},
  \href {https://ui.adsabs.harvard.edu/abs/2019ApJ...875...77P} {875, 77}

\bibitem[\protect\citeauthoryear{{Pawlowski}}{{Pawlowski}}{2018}]{Pawlowski2018}
{Pawlowski} M.~S.,  2018, \mn@doi [Modern Physics Letters A]
  {10.1142/S0217732318300045}, \href
  {https://ui.adsabs.harvard.edu/abs/2018MPLA...3330004P} {33, 1830004}

\bibitem[\protect\citeauthoryear{{Pawlowski} \& {Kroupa}}{{Pawlowski} \&
  {Kroupa}}{2020}]{Pawlowski2020}
{Pawlowski} M.~S.,  {Kroupa} P.,  2020, \mn@doi [\mnras]
  {10.1093/mnras/stz3163}, \href
  {https://ui.adsabs.harvard.edu/abs/2020MNRAS.491.3042P} {491, 3042}

\bibitem[\protect\citeauthoryear{{Pawlowski}, {McGaugh}  \&
  {Jerjen}}{{Pawlowski} et~al.}{2015}]{Pawlowski2015}
{Pawlowski} M.~S.,  {McGaugh} S.~S.,   {Jerjen} H.,  2015, \mn@doi [\mnras]
  {10.1093/mnras/stv1588}, \href
  {https://ui.adsabs.harvard.edu/abs/2015MNRAS.453.1047P} {453, 1047}

\bibitem[\protect\citeauthoryear{Pearson, K\"upper, Johnston  \&
  Price-Whelan}{Pearson et~al.}{2015}]{Pe15}
Pearson S.,  K\"upper A. H.~W.,  Johnston K.~V.,   Price-Whelan A.~M.,  2015,
  \mn@doi [\apj] {10.1088/0004-637x/799/1/28}, 799, 28

\bibitem[\protect\citeauthoryear{{Petersen} \& {Pe{\~n}arrubia}}{{Petersen} \&
  {Pe{\~n}arrubia}}{2020}]{Pe20}
{Petersen} M.~S.,  {Pe{\~n}arrubia} J.,  2020, \mn@doi [\mnras]
  {10.1093/mnrasl/slaa029}, \href
  {https://ui.adsabs.harvard.edu/abs/2020MNRAS.494L..11P} {494, L11}

\bibitem[\protect\citeauthoryear{{Petersen}, {Katz}  \& {Weinberg}}{{Petersen}
  et~al.}{2016a}]{PKW}
{Petersen} M.~S.,  {Katz} N.,   {Weinberg} M.~D.,  2016a, \mn@doi [\prd]
  {10.1103/PhysRevD.94.123013}, \href
  {https://ui.adsabs.harvard.edu/abs/2016PhRvD..94l3013P} {94, 123013}

\bibitem[\protect\citeauthoryear{{Petersen}, {Weinberg}  \& {Katz}}{{Petersen}
  et~al.}{2016b}]{Pe16}
{Petersen} M.~S.,  {Weinberg} M.~D.,   {Katz} N.,  2016b, \mn@doi [\mnras]
  {10.1093/mnras/stw2141}, \href
  {https://ui.adsabs.harvard.edu/abs/2016MNRAS.463.1952P} {463, 1952}

\bibitem[\protect\citeauthoryear{{Planck Collaboration (Ade et al.)}}{{Planck
  Collaboration (Ade et al.)}}{2014}]{planck}
{Planck Collaboration (Ade et al.)} 2014, \mn@doi [\aap]
  {10.1051/0004-6361/201321591}, \href
  {http://adsabs.harvard.edu/abs/2014A%26A...571A..16P} {571, A16}

\bibitem[\protect\citeauthoryear{{Polyachenko} \& {Shukhman}}{{Polyachenko} \&
  {Shukhman}}{1981}]{polyachenko1981}
{Polyachenko} V.~L.,  {Shukhman} I.~G.,  1981, \sovast, \href
  {http://adsabs.harvard.edu/abs/1981SvA....25..533P} {25, 533}

\bibitem[\protect\citeauthoryear{{Prada}, {Forero-Romero}, {Grand}, {Pakmor}
  \& {Springel}}{{Prada} et~al.}{2019}]{Pr19}
{Prada} J.,  {Forero-Romero} J.~E.,  {Grand} R. J.~J.,  {Pakmor} R.,
  {Springel} V.,  2019, \mn@doi [\mnras] {10.1093/mnras/stz2873}, \href
  {https://ui.adsabs.harvard.edu/abs/2019MNRAS.490.4877P} {490, 4877}

\bibitem[\protect\citeauthoryear{{Reid} \& {Brunthaler}}{{Reid} \&
  {Brunthaler}}{2004}]{ReidBrunthaler2004}
{Reid} M.~J.,  {Brunthaler} A.,  2004, \mn@doi [\apj] {10.1086/424960}, \href
  {https://ui.adsabs.harvard.edu/abs/2004ApJ...616..872R} {616, 872}

\bibitem[\protect\citeauthoryear{{Sanders} \& {Evans}}{{Sanders} \&
  {Evans}}{2017}]{Sa17}
{Sanders} J.~L.,  {Evans} N.~W.,  2017, \mn@doi [\mnras]
  {10.1093/mnras/stx2116}, \href
  {https://ui.adsabs.harvard.edu/abs/2017MNRAS.472.2670S} {472, 2670}

\bibitem[\protect\citeauthoryear{{Sanders}, {Evans}  \& {Dehnen}}{{Sanders}
  et~al.}{2018}]{SandersCrater}
{Sanders} J.~L.,  {Evans} N.~W.,   {Dehnen} W.,  2018, \mn@doi [\mnras]
  {10.1093/mnras/sty1278}, \href
  {https://ui.adsabs.harvard.edu/abs/2018MNRAS.478.3879S} {478, 3879}

\bibitem[\protect\citeauthoryear{{Sch{\"o}nrich}, {Binney}  \&
  {Dehnen}}{{Sch{\"o}nrich} et~al.}{2010}]{Schoenrich}
{Sch{\"o}nrich} R.,  {Binney} J.,   {Dehnen} W.,  2010, \mn@doi [\mnras]
  {10.1111/j.1365-2966.2010.16253.x}, \href
  {https://ui.adsabs.harvard.edu/abs/2010MNRAS.403.1829S} {403, 1829}

\bibitem[\protect\citeauthoryear{{Sellwood}}{{Sellwood}}{2003}]{Sellwood03}
{Sellwood} J.~A.,  2003, \mn@doi [\apj] {10.1086/368285}, \href
  {https://ui.adsabs.harvard.edu/abs/2003ApJ...587..638S} {587, 638}

\bibitem[\protect\citeauthoryear{{Shao}, {Cautun}, {Frenk}, {Grand },
  {G{\'o}mez}, {Marinacci}  \& {Simpson}}{{Shao} et~al.}{2018}]{Shao2018}
{Shao} S.,  {Cautun} M.,  {Frenk} C.~S.,  {Grand } R. J.~J.,  {G{\'o}mez}
  F.~A.,  {Marinacci} F.,   {Simpson} C.~M.,  2018, \mn@doi [\mnras]
  {10.1093/mnras/sty343}, \href
  {https://ui.adsabs.harvard.edu/abs/2018MNRAS.476.1796S} {476, 1796}

\bibitem[\protect\citeauthoryear{{Shao}, {Cautun}  \& {Frenk}}{{Shao}
  et~al.}{2019}]{Shao2019}
{Shao} S.,  {Cautun} M.,   {Frenk} C.~S.,  2019, \mn@doi [\mnras]
  {10.1093/mnras/stz1741}, \href
  {https://ui.adsabs.harvard.edu/abs/2019MNRAS.488.1166S} {488, 1166}

\bibitem[\protect\citeauthoryear{{Silverman}}{{Silverman}}{1986}]{Silverman}
{Silverman} R.,  1986, {Density Estimation}.
{Chapman and Hall, London}

\bibitem[\protect\citeauthoryear{{Simon} et~al.,}{{Simon}
  et~al.}{2020}]{Simon2020}
{Simon} J.~D.,  et~al., 2020, \mn@doi [\apj] {10.3847/1538-4357/ab7ccb}, \href
  {https://ui.adsabs.harvard.edu/abs/2020ApJ...892..137S} {892, 137}

\bibitem[\protect\citeauthoryear{{Siopis} et~al.,}{{Siopis}
  et~al.}{2009}]{Siopis09}
{Siopis} C.,  et~al., 2009, \mn@doi [\apj] {10.1088/0004-637X/693/1/946}, \href
  {https://ui.adsabs.harvard.edu/abs/2009ApJ...693..946S} {693, 946}

\bibitem[\protect\citeauthoryear{{Springel}}{{Springel}}{2005}]{springel_2005}
{Springel} V.,  2005, \mn@doi [\mnras] {10.1111/j.1365-2966.2005.09655.x},
  \href {http://adsabs.harvard.edu/abs/2005MNRAS.364.1105S} {364, 1105}

\bibitem[\protect\citeauthoryear{{Torrealba} et~al.,}{{Torrealba}
  et~al.}{2019}]{Torrealba2019}
{Torrealba} G.,  et~al., 2019, \mn@doi [\mnras] {10.1093/mnras/stz1624}, \href
  {https://ui.adsabs.harvard.edu/abs/2019MNRAS.488.2743T} {488, 2743}

\bibitem[\protect\citeauthoryear{{Valluri}, {Merritt}  \& {Emsellem}}{{Valluri}
  et~al.}{2004}]{ValluriME04}
{Valluri} M.,  {Merritt} D.,   {Emsellem} E.,  2004, \mn@doi [\apj]
  {10.1086/380896}, \href
  {https://ui.adsabs.harvard.edu/abs/2004ApJ...602...66V} {602, 66}

\bibitem[\protect\citeauthoryear{{Vasiliev}}{{Vasiliev}}{2013}]{Vasiliev2013}
{Vasiliev} E.,  2013, \mn@doi [\mnras] {10.1093/mnras/stt1235}, \href
  {http://adsabs.harvard.edu/abs/2013MNRAS.434.3174V} {434, 3174}

\bibitem[\protect\citeauthoryear{{Vasiliev}}{{Vasiliev}}{2018}]{Va18}
{Vasiliev} E.,  2018 (\mn@eprint {arxiv} {1802.08255})

\bibitem[\protect\citeauthoryear{{Vasiliev}}{{Vasiliev}}{2019}]{Va19}
{Vasiliev} E.,  2019, \mn@doi [\mnras] {10.1093/mnras/sty2672}, \href
  {https://ui.adsabs.harvard.edu/abs/2019MNRAS.482.1525V} {482, 1525}

\bibitem[\protect\citeauthoryear{{Vera-Ciro} \& {Helmi}}{{Vera-Ciro} \&
  {Helmi}}{2013}]{Ve13}
{Vera-Ciro} C.,  {Helmi} A.,  2013, \mn@doi [\apjl]
  {10.1088/2041-8205/773/1/L4}, \href
  {https://ui.adsabs.harvard.edu/abs/2013ApJ...773L...4V} {773, L4}

\bibitem[\protect\citeauthoryear{{Wang}, {Athanassoula}  \& {Mao}}{{Wang}
  et~al.}{2020}]{Wang20}
{Wang} Y.,  {Athanassoula} E.,   {Mao} S.,  2020, arXiv, \href
  {https://ui.adsabs.harvard.edu/abs/2020arXiv200504807W} {p. 2005.04807}

\bibitem[\protect\citeauthoryear{{Weinberg}}{{Weinberg}}{1999}]{weinberg1999}
{Weinberg} M.~D.,  1999, \mn@doi [\aj] {10.1086/300669}, \href
  {http://adsabs.harvard.edu/abs/1999AJ....117..629W} {117, 629}

\bibitem[\protect\citeauthoryear{{Zhao}}{{Zhao}}{1996}]{zhao1996}
{Zhao} H.,  1996, \mn@doi [\mnras] {10.1093/mnras/278.2.488}, \href
  {http://adsabs.harvard.edu/abs/1996MNRAS.278..488Z} {278, 488}

\bibitem[\protect\citeauthoryear{{de Lorenzi}, {Debattista}, {Gerhard}  \&
  {Sambhus}}{{de Lorenzi} et~al.}{2007}]{deLorenzi07}
{de Lorenzi} F.,  {Debattista} V.~P.,  {Gerhard} O.,   {Sambhus} N.,  2007,
  \mn@doi [\mnras] {10.1111/j.1365-2966.2007.11434.x}, \href
  {https://ui.adsabs.harvard.edu/abs/2007MNRAS.376...71D} {376, 71}

\makeatother
\end{thebibliography}

\appendix

\section{Generalised NFW models}\label{sec:nfw}

In this appendix, we give the expressions for the potential and density of this basis set, derived from \citet{Lilley2018b}.  At zeroth-order ($n=l=m=0$), this one-parameter family of basis functions exactly match the `generalised NFW' profiles studied in \citet{Ev06}.
All quantities can be expressed in terms of elementary functions. The angular variation is expanded using spherical harmonics $Y_{lm}$, and the radial terms are constructed out of the Jacobi polynomials. 

We write the basis functions in forms that obey the relations
\begin{equation}\label{eqn:basis_set}
  \int \dif r \: r^2 \: {\Phi}_{nl} {\rho}_{n^\prime l} = \delta_{n n^\prime} N_{nl},\quad
  \nabla^2\left({\Phi}_{nl}Y_{lm}\right) = 4\pi K_{nl}{\rho}_{nl}Y_{lm}.
\end{equation}
The zeroth-order potential is given by
\begin{equation}
  \Phi_{00} = %
  \begin{cases}
    {\displaystyle \frac{\log{(1+r)}}{r}},& \text{if } \nu = 0\\
    \null&\null\\
    {\displaystyle \frac{1 - (1+r)^{-\nu}}{\nu \: r}}, & \text{otherwise}.
  \end{cases}
\end{equation}
so that $\nu=0$ corresponds to the famous NFW model of \citet{NFW1997}.

At fixed $r$, higher-order terms in $l$ of the potential are given by the recurrence relation
\begin{equation}\label{eqn::phi_recursion}
\begin{split}
\Phi_{0,l+1} &= \frac{f_l}{r} \biggl\{\Phi_{0l}
   - \frac{r^l}{(1+r)^{1+2l+\nu}} \left[\frac{1+2l+\nu}{2+2l}\:\frac{r}{1+r} + 1\right]\biggr\},\\
f_l &= \frac{(3+2l)(2+2l)}{(1+2l+\nu)(2+2l+\nu)}.
\end{split}
\end{equation}
Note that, while formally correct, the recurrence relation \eqref{eqn::phi_recursion} suffers from catastrophic cancellation when $l$ is high and $r$ is low. In these situations it is therefore more accurate to use a few terms of the following Taylor expansion
\begin{equation}\label{eqn::phi_expansion}
\begin{split}
  \Phi_{0l} \approx \frac{r^l}{(1+r)^{2l+\nu}}\left[1 - \frac{(1-\nu)}{2+2l}r + \frac{(1-\nu)(2-\nu)}{(2+2l)(3+2l)}r^2\right.\\
  \left. \ldots + \frac{(1-\nu)\ldots(j-\nu)}{(2+2l)\ldots(j+1+2l)}(-r)^j\right].
\end{split}
\end{equation}
A suitable algorithm to compute $\Phi_{0l}$ to at least 6 digits of accuracy over the entire parameter space covered in this paper would be
\begin{equation*}
  \Phi_{0l} = %
  \begin{cases}
    \text{if } r<10^{-4/(l+1)}, \text{use \eqref{eqn::phi_expansion} keeping terms up to } j=4,\\
    \text{otherwise use \eqref{eqn::phi_recursion}}.
  \end{cases}
\end{equation*}
Higher-order terms in $n$ of the potential are given by the recurrence relation
\begin{equation}
\Phi_{n+1,l} = \Phi_{nl} - \frac{2n!}{(2+2l)_n} \: \frac{r^l}{(1+r)^{1+2l+\nu}} \: P_n^{(2l+2\nu,2l+1)}(\xi),
\end{equation}
where $P_n^{(\alpha,\beta)}(x)$ are the Jacobi polynomials and $\xi = (r-1)/(r+1)$. Similarly, the radial component of acceleration is given by
\begin{equation}
\begin{split}
  \Phi_{nl}^\prime = - \frac{(1+l)\Phi_{nl}}{r} + \frac{A_{nl}r^{l-1}}{(1+r)^{1+2l+\nu}}
  \Big[&(n\!+\!4l\!+\!2\nu\!+\!1)P_n^{(2l+2\nu,2l+1)}(\xi)
   \\&- (n\!+\!2l\!+\!2\nu) P_{n-1}^{(2l+2\nu,2l+1)}(\xi)
   \Big],
\end{split}
\end{equation}
where $A_{nl} \equiv \frac{1+2l}{2n+4l+2\nu+1}$; and the density functions are given simply by
\begin{equation}
\begin{split}
  \rho_{nl} = \frac{r^{l-1}}{(1+r)^{2+2l+\nu}}
  \Big[&(n\!+\!4l\!+\!2\nu\!+\!1)(n\!+\!2l\!+\!\nu\!+\!1)P_n^{(2l+2\nu,2l+1)}(\xi)
   \\&- (n\!+\!2l\!+\!2\nu)(n\!+\!2l\!+\!\nu) P_{n-1}^{(2l+2\nu,2l+1)}(\xi)
   \Big].
\end{split}
\end{equation}
In this way, the potential, acceleration and density functions may be constructed (for a given $l$) from a single ladder of recursively-computed Jacobi polynomials $P_{n}^{(2l+2\nu,2l+1)}(\xi)$. In principle, one could find a recurrence relation that connected basis functions at adjacent values of $l$, but the present scheme (recursing first in $l$ for the $n=0$ functions, and subsequently recursing in $n$) is already optimal, with $nl$ operations in total.

The associated constants $N_{nl}$ and $K_{nl}$ are
\begin{equation}
\begin{split}
N_{nl} &= \frac{(2l+1)!}{(n+2l+2\nu+1)_{2l}},\\
K_{nl} &= -\frac{n!(2l+1)}{4\pi (2n+4l+2\nu+1) (2l+1)_n}.
\end{split}
\end{equation}
For these models the inner slope is fixed at $\gamma=1$, and the parameter $\nu$ adjusts the outer slope $\beta$, so we could alternatively use this as the free parameter, writing $\beta = 3 + \nu$.
A similar basis set, with fixed outer slope $\beta = 4$ and variable inner slope $\gamma = 2-\nu$, can be obtained by the transformations $\Phi_{nl}(r) \mapsto r^{-1}\Phi_{nl}(r^{-1})$ and $\rho_{nl}(r) \mapsto r^{-5}\rho_{nl}(r^{-1})$. Further discussion of this and a more general family of basis sets may be found in \citet{Li20}.

\section{Derivation of tail coefficients}\label{sec:tail_coefficients}

A hard truncation in the particle distribution gives rise to underestimates of the acceleration at low order and ringing at high order for the biorthonormal potential expansion. To avoid these undesirable features, the density distribution is extrapolated beyond the cutoff assuming it follows a power law. In this appendix, we give expressions for the calculation of the modified coefficients.

Our N-body halo consists of $N$ particles with masses $m_j$ at positions $\boldsymbol{r}_j$ (with $r_j \leq \rt$). For the region $\rt < r < \infty$, we affix to the N-body data an analytical `tail' density corresponding to the underlying zeroth-order density model of our chosen basis set:
\begin{equation}\label{eqn:modified_rho_hat}
  \hat\rho(\boldsymbol{r}) =
  \begin{cases}
    \sum_j m_j \delta^3(\boldsymbol{r} - \boldsymbol{r}_j), & \mathrm{if} \: r \leq \rt \\
    \mathcal{A} \rho_{000}(r), & \mathrm{if} \: r > \rt,
    \end{cases}
\end{equation}
where $\mathcal{A}$ is a constant that sets the normalisation of this `tail' profile. Because Eq.~(\ref{eqn:modified_rho_hat}) defines a linear adjustment to the data, the coefficients of the basis expansion can now simply be linearly corrected to take into account this \lq tail' density. We denote the original \lq uncorrected' coefficients by $C^\mathrm{orig}_{nlm}$, and the coefficients corresponding to this \lq tail' density by $T_{nlm}$: 
\begin{equation}
\begin{split}
  C^\mathrm{orig}_{nlm}  &\equiv \sum_i m_i \: \Phi_{nlm}(\boldsymbol{r}_i),\\
  T_{nlm}  &\equiv \int_{r>\rt} \Phi_{nlm} \: \rho_{000} \dddr,\\
  C_{nlm}  &\equiv C^\mathrm{orig}_{nlm} + \mathcal{A} T_{nlm}.
 \end{split}
\end{equation}
In order to fix the parameter $\mathcal{A}$, we pick a radius $R$ and constrain the expansion to have the same mass interior to $R$ as the N-body halo -- this could be any radius, but in practice we use $R=\rt$. Denoting by $M_{nlm}(R)$ the mass enclosed at radius $R$ by the $n$-th basis function (an analytical quantity that is non-zero only when $m=l=0$), we have:
\begin{equation}
M_{nlm}(R) \equiv \int_{r<R} \rho_{nlm} \dddr = \delta_{l0} \delta_{m0} R^2 \left.\frac{\dif\Phi_{n0}(r)}{\dif r}\right|_{r=R}.
\end{equation}
Then let $M_\mathrm{enc}^\mathrm{true}(R)$ and $M_\mathrm{enc}^\mathrm{orig}(R)$ be the mass interior to $R$ of the corrected and uncorrected expansions, respectively:
\begin{equation}\label{eqn:Menc_expansion}
\begin{split}
  M_\mathrm{enc}^\mathrm{true}(R) &\equiv \int_{r<R} \rho(\boldsymbol{r}) \dddr = \sum_{n} C_{n00} \: M_{n00}(R),\\
  M_\mathrm{enc}^\mathrm{orig}(R) &\equiv \int_{r<R} \rho^\mathrm{orig}(\boldsymbol{r}) \dddr = \sum_{n} C^\mathrm{orig}_{n00} \: M_{n00}(R).
  \end{split}
\end{equation}
Let $M_\mathrm{enc}^\mathrm{tail}(R)$ be the (unnormalised) mass of the \lq tail' portion of the density profile
\begin{equation}\label{eqn:Menc_tail}
  M_\mathrm{enc}^\mathrm{tail}(R) \equiv \sum_{n} T_{n00}(\rt) \: M_{n00}(R),
\end{equation}
so we clearly have
\begin{equation}
\begin{split}
  M_\mathrm{enc}^\mathrm{true}(R) = \sum_{n} \left[C^\mathrm{orig}_{n00} + \mathcal{A} T_{n00}\right] M_{n00}(R),\\ = M_\mathrm{enc}^\mathrm{orig}(R) + \mathcal{A} M_\mathrm{enc}^\mathrm{tail}(R).
\end{split}
\end{equation}
And finally let $M_\mathrm{enc}^\mathrm{N-body}(R)$ be the mass of the total particles in the halo that are interior to $R$,
\begin{equation}
  M_\mathrm{enc}^\mathrm{N-body}(R) \equiv \sum_{r_j<R} m_j.
\end{equation}
Then to fix the value of $\mathcal{A}$, we simply require that $M_\mathrm{enc}^\mathrm{true}(R) = M_\mathrm{enc}^\mathrm{N-body}(R)$, giving
\begin{equation}
  \mathcal{A} = \frac{M_\mathrm{enc}^\mathrm{true}(R) - M_\mathrm{enc}^\mathrm{orig}(R)}{M_\mathrm{enc}^\mathrm{tail}(R)}.
\end{equation}

The quantities $T_{nlm}$ can be calculated in advance. We require the integral over the interval $(R,\infty)$ between the zeroth-order density and each potential basis function, which is non-zero only when $m=l=0$. Assuming unit-normalised spherical harmonics and working in units where $G = 1$ and the scale-length $\rs = 1$, we have the result
\begin{equation}\label{eqn:Tnlm}
\begin{split}
  T_{n00}(R)
  &= \int_{R}^\infty r^2 \dif r \: \Phi_{n0}\rho_{00} \\
  &=
  \delta_{n0}N_{00} - \frac{1}{4\pi K_{00}}R^2\left(\frac{\dif \Phi_{00}}{\dif r}\Phi_{n0}-
    \frac{\dif\Phi_{n0}}{\dif r}\Phi_{00}\right)_{r=R} 
    \\& - \frac{K_{n0}}{K_{00}} \int_0^{R} r^2 \dif r \: \Phi_{00} \rho_{n0}.
\end{split}
\end{equation}
%
%
%
It remains to evaluate the last integral in Eq.~\eqref{eqn:Tnlm}, which is
%
\begin{equation}
\begin{split}
\int_0^{R} r^2 \dif r \: \Phi_{00} \rho_{n0} &= (n+1) \: I_n\!\left(\chi\right) - n \: I_{n-1}\!\left(\chi\right),
\end{split}
\end{equation}
where we have defined $\chi \equiv R/(1 + R)$, and
%
\begin{equation}\label{eqn:define_chi_In_Qj}
\begin{split}
  I_n(\chi) &\equiv (n\!+\!\nu\!+\!1)(n\!+\!2\nu\!+\!1)\sum_{j=0}^n\frac{(-1)^{n-j} (n\!+\!2\nu\!+\!2)_j}{(j\!+\!1)!} 
  \binom{n}{j}Q_j(\chi),\\
  Q_j(\chi) &\equiv
  \begin{cases}
    {\displaystyle \frac{1}{1+j}\left[\chi^{1+j}\left(\frac{1}{1+j}-\log{(1-\chi)}\right) - \Beta_\chi(1+j,0)\right]},& \null \\
    \qquad\qquad\qquad\qquad\qquad\qquad\qquad\text{if } \nu = 0, &\null\\
    \null&\null\\
    {\displaystyle \frac{1}{\nu}\left[\Beta_\chi(1+j,1+\nu) - \Beta_\chi(1+j,1+2\nu)\right]}, & \null\\
    \qquad\qquad\qquad\qquad\qquad\qquad\qquad \text{otherwise.}&\null
  \end{cases}
\end{split}
\end{equation}
Here, $\Beta_z(a,b)$ is the incomplete beta function and we have made use of the Pochhammer symbol $(z)_j$ to indicate the falling factorial.

The formulae in this Appendix hold good for the generalised NFW basis set used in the main body of this paper. Analogous formulae that cover the full parameter space of possible basis sets may be found in \citet{Li20}.

\bsp	
\label{lastpage}
\end{document}